\documentclass[sigconf]{acmart}

\AtBeginDocument{%
  }

\copyrightyear{2025}
\acmYear{2025}
\setcopyright{acmlicensed}\acmConference[DIS '25]{Designing Interactive
Systems Conference}{July 5--9, 2025}{Funchal, Portugal}
\acmBooktitle{Designing Interactive Systems Conference (DIS '25), July 5--9,
2025, Funchal, Portugal}
\acmDOI{10.1145/3715336.3735685}
\acmISBN{979-8-4007-1485-6/2025/07}

\pdfoutput=1
\usepackage{subfigure}
\usepackage{subcaption}
\usepackage{xcolor}

\definecolor{myblue}{RGB}{244, 246, 253}

\begin{document}

%\title{On Demand Audio Descriptions for Blind and Low Vision Users}
%\title[On demand AD]{On Demand Audio Descriptions for Blind and Low Vision Users}
\title{Describe Now: User-Driven Audio Description for Blind and Low Vision Individuals}
%On-Demand Audio Description to Enhance Accessibility of Online Videos for Blind and Low Vision Users} 
%On-Demand Audio Description to Enhance Access and Control of Online Videos for Blind and Low Vision Users
%On-Demand Audio Description: Enhancing Video Accessibility and Control for Blind and Low Vision Users
%On-Demand Audio Description to Enhance Blind and Low Vision Users' Control of Video Accessibility

%
% The "author" command and its associated commands are used to define
% the authors and their affiliations.
% Of note is the shared affiliation of the first two authors, and the
% "authornote" and "authornotemark" commands
% used to denote shared contribution to the research.
% \author{Anonymous Author(s)}
% \author{Maryam Cheema}
% \email{mcheema2@asu.edu}
% \orcid{0009-0009-0556-8029}
% \author{Hasti Seifi}
% \email{Hasti.Seifi@asu.edu}
% \orcid{0000-0001-6437-0463}
% \author{Pooyan Fazli}
% \email{pooyan@asu.edu}
% \affiliation{%
%   \institution{Arizona State University}
%   \city{Tempe}
%   \state{Arizona}

\author{Maryam Cheema}
\affiliation{%
  % \department{School of Computing and Augmented Intelligence}
  \institution{Arizona State University}
  \city{Tempe}
  \state{Arizona}
  \country{USA}
}
\email{mcheema2@asu.edu}
\author{Hasti Seifi}
\authornote{Hasti Seifi and Pooyan Fazli are the corresponding authors.}
\affiliation{%
  % \department{School of Computing and Augmented Intelligence}
  \institution{Arizona State University}
  \city{Tempe}
  \state{Arizona}
  \country{USA}
}
\email{hasti.seifi@asu.edu}
\author{Pooyan Fazli}
\authornotemark[1]
\affiliation{%
%   \department{School of Arts, Media and
% Engineering}
  \institution{Arizona State University}
  \city{Tempe}
  \state{Arizona}
  \country{USA}
}
\email{pooyan@asu.edu}

% \renewcommand{\shortauthors}{Cheema et al.}

\begin{teaserfigure}
\centering
\includegraphics[width=\textwidth]{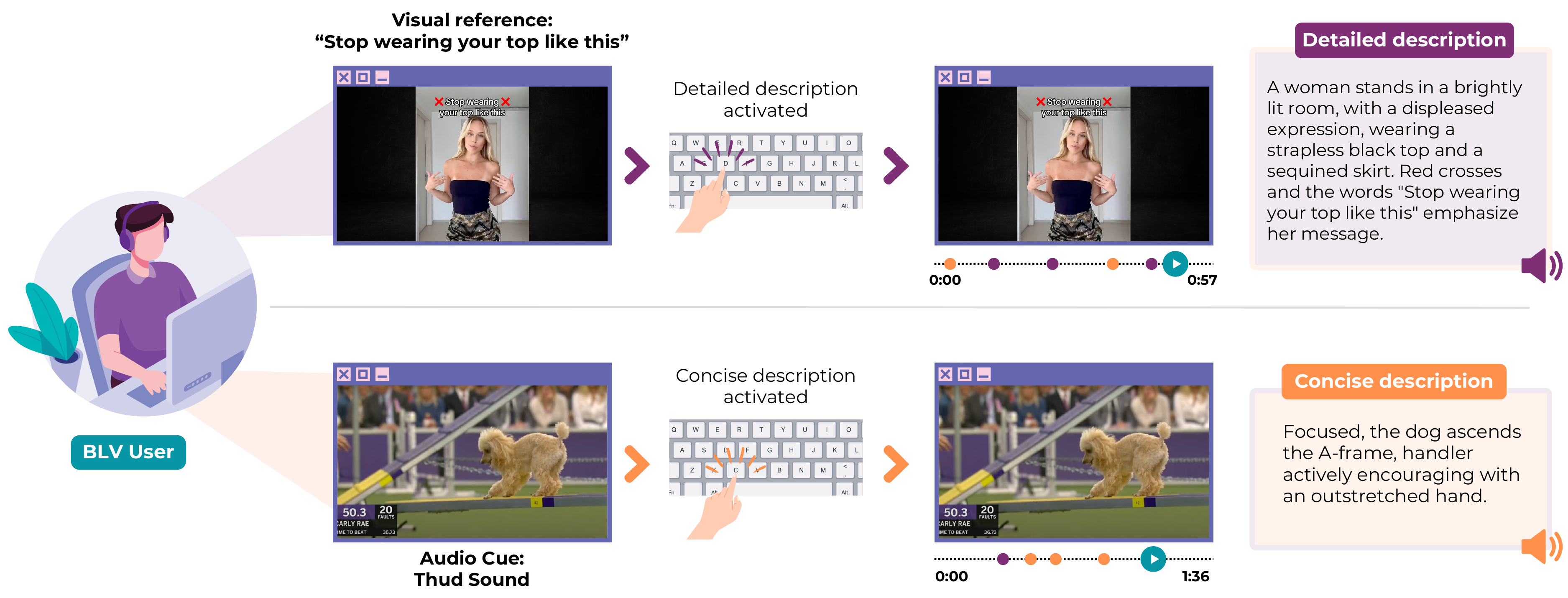}
\caption{User-driven descriptions for two video genres. When watching a video, blind and low vision users can press the C or D keys to activate concise and detailed descriptions generated by a multi-modal large language model. In the user study, participants activated descriptions based on different use cases such as visual references or audio cues in the main audio track.} % The descriptions are generated using a large language model (GPT-4-Vision).}
\Description{Illustration depicting user-driven audio descriptions for different video genres. At the top, a BLV (blind or low vision) user interacts with a beauty video showing a woman with the text 'Stop wearing your top like this.' The user presses the “D” key to activate a 'Detailed description,' which describes the woman’s displeased expression and outfit (a strapless black top and sequined skirt), with red crosses emphasizing her message. The figure illustrates how the visual reference of the top is used to activate the detailed description. At the bottom, another video (Pets & Animals) shows a dog ascending an A-frame in an agility course. The BLV user presses a key to activate a 'Concise description,' which describes the dog’s actions and the handler’s encouragement. Audio cues, such as a 'Thud Sound,' are mentioned for activating the concise descriptions, and a timeline bar shows when descriptions are triggered.}
\label{fig:teaser}
\end{teaserfigure}

\begin{abstract}
 Audio descriptions (AD) make videos accessible for blind and low vision (BLV) users by describing visual elements that cannot be understood from the main audio track. AD created by professionals or novice describers is time-consuming and offers little customization or control to BLV viewers on description length and content and when they receive it. To address this gap, we explore user-driven AI-generated descriptions, enabling BLV viewers to control both the timing and level of detail of the descriptions they receive. In a study, 20 BLV participants activated audio descriptions for seven different video genres with two levels of detail: concise and detailed. %Our user study delves into on-demand descriptions' effectiveness, efficiency, and enjoyability through testing with 20 BLV participants. 
 %We report the frequency and detail level of on-demand requests for different videos and user ratings of effectiveness, efficiency and enjoyability of watching videos with on-demand ADs, and users.  %
Our findings reveal differences in the preferred frequency and level of detail of ADs for different videos, participants' sense of control with this style of AD delivery, and its limitations. We discuss the implications of these findings for the development of future AD tools for BLV users.
 
%Our results show differences in AD frequency and level of detail BLV users wanted for different videos, 
 %Findings reveal insights into preferences for concise AD for instructional content and detailed AD for entertainment content. %Results also highlight statistical differences in the frequency of requiring descriptions for different video genres. 
 %their sense of control with this style of AD delivery, its limitations, and variations among BLV users in their perception of user-driven AI descriptions. We discuss the implications of our findings for future AD tools for BLV users.
\end{abstract}





% \received{13th January 2025}
% \received[revised]{}
% \received[accepted]{}

\maketitle

\section{Introduction}
The rapid growth and popularity of videos on online platforms have widened the accessibility gap for blind and low vision (BLV) users. With billions of users using platforms such as YouTube, TikTok, and Instagram, videos make up 82\% of all internet traffic~\cite{synthesia_video_statistics, hubspot_video_consumption}. As a result, video content has become more diverse, with a wide range of user-generated content and varying levels of quality. To make videos accessible, professional describers record audio descriptions (AD) to narrate the key visual elements, such as actions, characters, scene changes, on-screen text, and other relevant content~\cite{visual_made_verbal}. While community platforms, such as YouDescribe~\cite{youdescribe_website}, allow sighted describers to volunteer and describe video content, %novice describers can take almost 1.5 hours to describe a 3-minute video on a topic they are familiar with~\cite{you_described_we_archived}.
%Thus, most requested videos by BLV users remain undescribed. In other words,
authoring pre-recorded ADs for the ever-growing number of online videos is simply not an option anymore.

Recent advances in artificial intelligence (AI) models, particularly multimodal large language models (MLLMs), provide automated methods and interfaces for generating descriptions~\cite{shortscribe,spica, wang_auto_ad, automated_AD,cineAD,chuang2023clearvid,nguyen2024oscar, chang2024worldscribe}. 
MLLMs can analyze a sequence of frames from a video and generate a textual description of the  content~\cite{LLAVA, openai_gptv_system_card, anthropic_claude, deepmind_gemini}. However, the effectiveness of AI-generated descriptions in directly serving BLV users remains largely unknown. Recent qualitative studies suggest that BLV users have diverse needs for the frequency and content of ADs depending on the type of video (e.g., entertainment vs.\ educational videos) and personal preferences~\cite{jiang2024s, wang_auto_ad}. Still, human and automated systems generally produce fixed descriptions at set times, which might not account for a BLV viewer's preference. %Furthermore, while automated systems can detect scene changes, they struggle with short videos, particularly those that are single-shot or feature rapid scene transitions~\cite{video_scene_detection}.
% which can be problematic for online videos with various lighting and video recording conditions. 
%While some guidelines for AD timing exist (e.g., \emph{``Avoid describing over audio that is essential to comprehension.''}~\cite{dcmp_description_key}), 
%Professional describers often rely on their tacit knowledge and the video content to time ADs in a way that supports BLV users' comprehension and enjoyment. 
%Furthermore, little data is available on how frequently ADs are needed for different types of videos. 
%Therefore, more research is needed to inform the design of AI-based AD platforms to serve the needs and preferences of BLV individuals for different types of videos.

To address these gaps, we investigate \emph{user-driven AI-generated descriptions} to analyze BLV users' perceptions of watching videos using this approach and identify their AD needs for different types of videos. User-driven ADs allow BLV viewers to activate a description at any point in the video based on their preferences and audio cues. While AI-generated descriptions can be verbose and inefficient for video consumption~\cite{verbositybiaspreferencelabeling}, detailed descriptions may enhance BLV users' understanding of visual content. Thus, we examine two levels of detail for ADs: a shorter, concise description vs.\ a more detailed description. To understand the utility of this approach, we ask: (Q1) What are BLV individuals' perceptions and experiences with user-driven AI-generated ADs? (Q2) How do BLV users' preferences for AD timing and detail differ between different video genres? 

To answer these questions, we developed a prototype for activating ADs and conducted a study with 20 BLV users across different video genres. We prompted an MLLM, specifically GPT-4 Vision (GPT-4V)~\cite{openai_gptv_system_card}, with AD guidelines from professional describers to generate descriptions for seven short online videos spanning various categories, such as \emph{film and animation}, \emph{Education}, and \emph{Cooking}.
% including film and animation, fashion and beauty, science education, food and cooking, health and fitness, people and blogs, and pets and animals. 
 % Then, we developed an interactive interface where 
 BLV users could press a key to activate either a concise or detailed description at any time. They interacted with the videos to activate ADs, rated the effectiveness, efficiency, and enjoyment of the ADs, and shared their experience with user-driven AI descriptions. 
%We analyzed user ratings and frequency of on-demand requests and applied open-coding and thematic analysis to the transcribed interviews. 
%Our quantitative results highlight significant differences in the frequency of descriptions needed for different video genres, with shorter AD intervals required for \emph{Film and Animation}, and longer intervals for \emph{Education, Health and Fitness}, and \emph{Beauty} videos. Moreover, BLV individuals differed in their frequency and type of AD requests. 
Our thematic analysis of the interviews identified three key themes: (1) BLV users' increased sense of control and active watching experience when activating ADs, accompanied by a higher cognitive workload, (2) preferences for pre-recorded vs.\ user-driven ADs depending on the video content, viewing context, and individual BLV users, and (3) the positive and negative aspects of user-driven AI descriptions. 
In addition, our quantitative results suggest differences in the frequency of descriptions needed for different video genres, with shorter AD intervals required for \emph{Film and Animation}, and longer intervals for \emph{Education} and \emph{Health and Fitness} videos. 
Drawing on these results, we discuss implications for future user-driven AD platforms and the evolving roles of BLV users and sighted describers. Our contributions include:

\begin{itemize}
%\item On-demand audio descriptions as a new way of watching short online videos and collecting data on BLV users' needs
\item Insights into BLV individuals' perceptions of user-driven AI descriptions when watching short online videos 
\item Empirical data on the frequency and type of AD requests for seven video genres as well as variations among BLV users
\end{itemize}

\section{Related Work}
We review prior work on video accessibility practices and needs and interactive tools for the creation and use of ADs.

\subsection{Audio Description Practices and Needs}
%Videos are generally inaccessible to BLV viewers, especially when the content is not discernable via audio alone. Hence video accessibility involves a range of practices that aim at making visual content accessible via auditory means. The most common approach is to add audio descriptions; they provide information about actions, characters, scene changes and on-screen text that are not described in the soundtrack of the video [7]. 
Although AD has been around for over three decades and produced for high-budget films and movies~\cite{visual_made_verbal,adaspedagolicaltool}, the field has gained momentum in the past decade due to the exponential amount of online videos~\cite{mazur2021audio}. %Traditionally, descriptions were created and pre-recorded by trained professionals for high-budget film and movies~\cite{visual_made_verbal}. However, the proliferation of online video content has led accessibility advocates and researchers  to emphasize the importance of meeting 
The WCAG 2.0 Level AA compliance mandates that AD be provided for all prerecorded video content in synchronized media~\cite{wcag_audio_desc, acb_audio_description}. This push for video accessibility has resulted in AD authoring guidelines~\cite{3play_audio_description, dcmp_description_key, netflix_style_guide, ofcom_access_services}, such as Netflix, which focuses on entertainment content, and DCMP, focusing on educational/instructional content. These guidelines, originally designed for professional describers, have been used to train novice describers on community-driven platforms~\cite{youdescribe_website}. 
% With the proliferation of online videos, accessibility advocates and researchers have developed AD guidelines and best practices for novice describers of video content. For example, Netflix includes XXXX and DCMP has XXXX. 
Yet, these guidelines and practices primarily focus on pre-recorded ADs. 
Furthermore, UK Ofcom guidelines state that ``some programmes are too fast-moving, or offer little opportunity to insert AD''~\cite{comparativestudyAD}, indicating a gap in how to describe other forms of video content.

Recent research suggested that BLV people wish to interact with video content in new ways besides pre-recorded ADs. Specifically, Bodi et al.~\cite{automated_AD} investigated the viability of providing video accessibility via interactive visual question answering and showed that BLV users requested descriptions more frequently than asking questions. Previous work investigating AD for 360$^\circ$ videos found that BLV participants wanted different parts of the visual scene described in addition to the main action~\cite{fidyka2018audio}. %had diverse preferences for the spatial location of AD~\cite{jiang360videos}. 
Other studies demonstrated that BLV participants
% , by creating spatial audio, scene and object descriptions. 
%BLV participants could listen to spatial ADs, feel vibrations for scene transitions, and explore objects. Study 
preferred immersive ADs over standard ADs for this format~\cite{omniscribe} and had diverse preferences for the spatial location of AD~\cite{jiang360videos}. Similarly, the SPICA system~\cite{spica} enabled BLV users to explore video content interactively. Participants found the object exploration feature and object-specific sound effects enhanced overall video consumption. These works highlight the importance of moving beyond traditional ADs to offer BLV users  %more user-driven interaction models, where receiving AD is 
a customized experience.  %To cater to this, we provided two levels of detail, shorter more concise descriptions, and longer more detailed descriptions. 
Similarly, our work investigates user-driven descriptions as a way for BLV users to access video content when needed. %, providing data on user perception and behavior of on-demand ADs across different video genres.

Others studied variations in BLV users' AD needs depending on their visual impairment and the video content.
Chmiel and Mazur~\cite{mazur_ad_preferences} examined AD preferences between congenitally blind, non-congenitally blind and low vision participants and %, on aspects such as character naming amd describing facial expressions and colors, using metaphors, explicitation and intertextual allusions. 
%Although difference in AD preferences weren't significant, there were 
found some specific preferences related to character naming and the use of metaphors in AD. %The study concludes AD that covers criteria for a broad range of participants should be provided, with the possibility of alternative version for specific groups. 
Another study highlighted how the amount of information in AD can impact the experience of BLV users~\cite{fresno}, where segmented stepwise descriptions were less cognitively demanding and led to better recall for BLV users. A recent interview study by Jiang et al.\ suggested that BLV users have different goals and preferences when watching different video genres~\cite{jiang2024s}. For example, BLV users wanted detailed descriptions of people and their appearance in short-form videos of family and friends but not in educational videos.  Relatedly, Natalie et al.\ found that customized ADs can improve BLV users' video understanding, immersion, and information navigation efficiency~\cite{natalie2024audio}. We build on these studies to collect quantitative data on the frequency and amount of detail needed for %Based on these findings, we included 
seven video genres %with concise and detailed ADs to evaluate the use of user-driven ADs 
across users with various visual impairments.

\vspace{-0.13cm}
\subsection{Interactive Systems for Audio Description}
%The explosion of online videos in the last decade has prompted the development of online platforms and interactive tools for audio description.  %Creating ADs is expensive and time-consuming. 
%YouDescribe is a community driven platform that enables sighted volunteers to create pre-recorded AD for YouTube videos. %BLV users can request audio descriptions for a video by adding it in their wish list to be described by a sighted volunteer. While YouDescribe expands on video accessibility by access to audio description creation, 
%Preparing audio descriptions is time-consuming and challenging. Reportedly, for novice describers, it takes almost 1.5 hours to describe a 3-minute video on a topic they are familiar with~\cite{you_described_we_archived}.
%Thus, most requested videos by BLV users on community platforms such as YouDescribe remain undescribed. 

To assist sighted users in creating descriptions, several tools have been built to streamline different parts of the process~\cite{live_describe,rescribe,crossa11y,kobayashi,yuksel2020HIML,viscene,accessiblead,natalie_feedback, pausesAD}. LiveDescribe was one of the first tools that investigated the potential of using volunteers to create audio descriptions by automating gaps to fit the audio description~\cite{live_describe}. Rescribe helped describers optimize description placement by dynamic programming~\cite{rescribe}, whereas CrossA11y assisted authors in detecting visual and auditory accessibility issues in a video~\cite{crossa11y}. 
% Kobayashi et al. created a tool that let users create descriptions that are converted from text to speech ~\cite{kobayashi}. 
Yuksel et al. developed a system that generated baseline descriptions which could then be revised by sighted individuals to produce high-quality descriptions~\cite{yuksel2020HIML}. Other tools have included BLV users in the AD and video creation process. 
% While not an AD creation platform, AVscript enables BLV content creators to edit videos via an accessible text based video editor, reducing the cognitive load of video editing~\cite{avscript}. 
Viscene studied the efficacy of collaboration between novice describers and BLV reviewers to create high-quality descriptions~\cite{viscene}. This work informed an automatic feedback tool to support novice describers in authoring ADs~\cite{natalie_feedback}. Similarly, AccessibleAD expanded access to AD writing to make BLV users become an active part of AD creation~\cite{accessiblead}. These approaches still require manual authorship, which even with technology support, remains a challenging task for describers.

Recent advances in artificial intelligence (AI) models have fueled research in automated AD authorship~\cite{accessibilityresearch}. Wang et al.\ built a system that analyzed the audio and visual content of the video to create automated descriptions using deep learning~\cite{wang_auto_ad}. Ihorn et al.\ developed a hybrid approach that generated descriptions and provided answers or additional information based on user queries~\cite{ihorn2021narrationbot}. Automated AD approaches have been dominated by deep learning, which focuses on visual content extraction (encoding) and text generation (decoding)~\cite{vdsurvey}. Due to the lack of context awareness, the automated descriptions can be rather verbose and or lack detail appropriate to the content of the video. 

To overcome these limitations, recent systems have used large language models (LLMs) to provide BLV users with agency when watching a video or scene~\cite{shortscribe,spica,chang2024worldscribe,li2025videoa11y}. For example, ShortScribe system utilized GPT-4 to provide three levels of hierarchical visual summaries
% provide long descriptions, short descriptions, and shot-by-shot descriptions 
for short-form videos. 
%This was the first work to provide different types of AI descriptions for videos. 
Similarly, SPICA used GPT-4 to create natural language descriptions of objects BLV users selected in key frames of a video~\cite{spica}. VideoA11y combined MLLMs with AD guidelines to generate descriptions that matched trained human annotations in clarity, accuracy, objectivity,
descriptiveness, and user satisfaction~\cite{li2025videoa11y}. WorldScribe further supported BLV user's agency by providing live visual descriptions about the user's surroundings using various vision language models~\cite{chang2024worldscribe}. %These systems have demonstrated the potential of user-driven approaches to descriptions. 
Our work builds on this prior research by using an MLLM to generate descriptions that can be activated at any time, to gather insights into how end-user control over description brevity and timing of descriptions will impact BLV users’ experiences.

\section{Study Materials and Interface} %STUDY MATERIALS AND INTERFACE}
To study user-driven AI descriptions, we selected short videos from different genres, created concise and detailed descriptions for them using GPT-4V, and developed a user interface for watching videos with user-driven ADs.
%Our study focuses on a selection of videos from different genres. The videos were audio described using GPT-4-Vision based on a set of guidelines established via guideline collection. 

\subsection{Video Selection}
\begin{figure}[htbp]
    \centering
   
        \subfigure[Test Video]{
        \includegraphics[width=4.0cm]{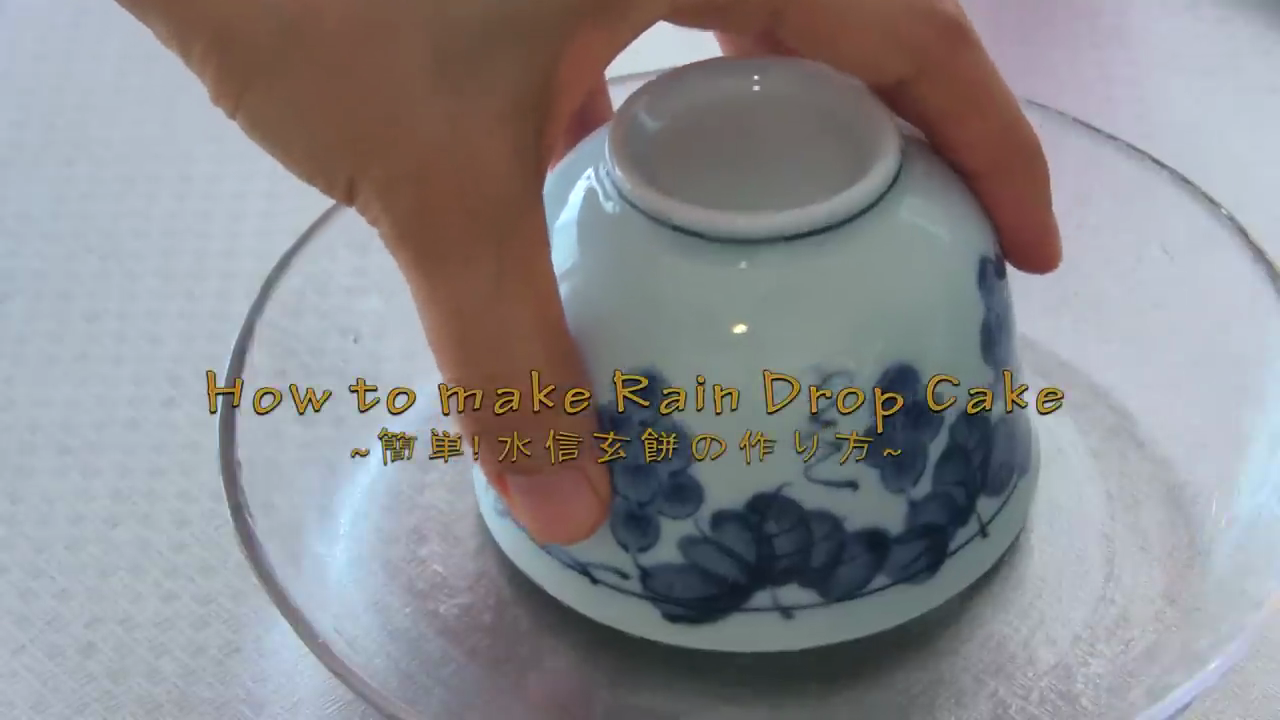}
        \label{fig:text}
    }
    \subfigure[People and Blogs Video]{
        \includegraphics[width=4.0cm]{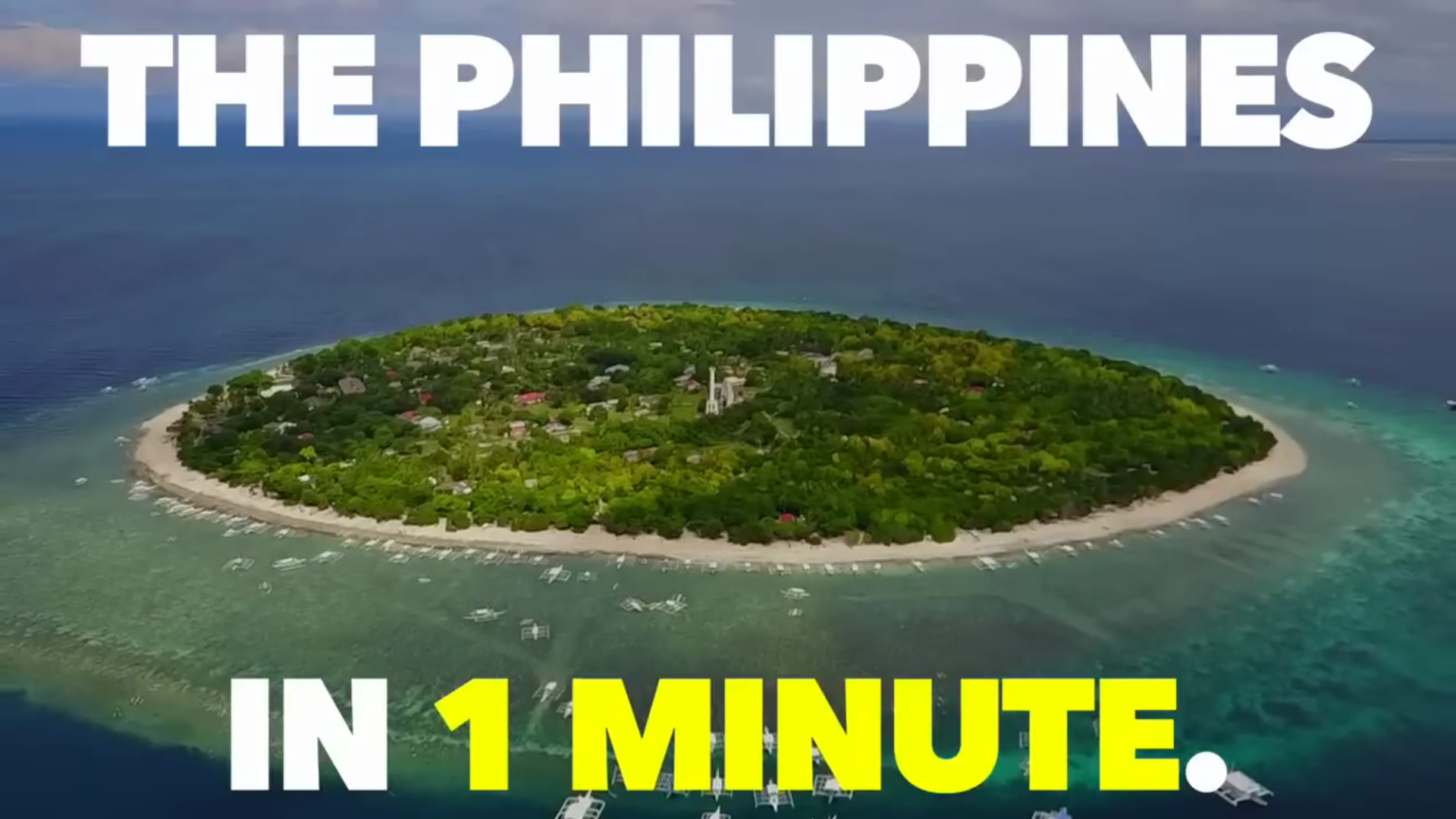}
        \label{fig:blog}
    }
    \subfigure[Pets \& Animals Video]{
        \includegraphics[width=4.0cm]{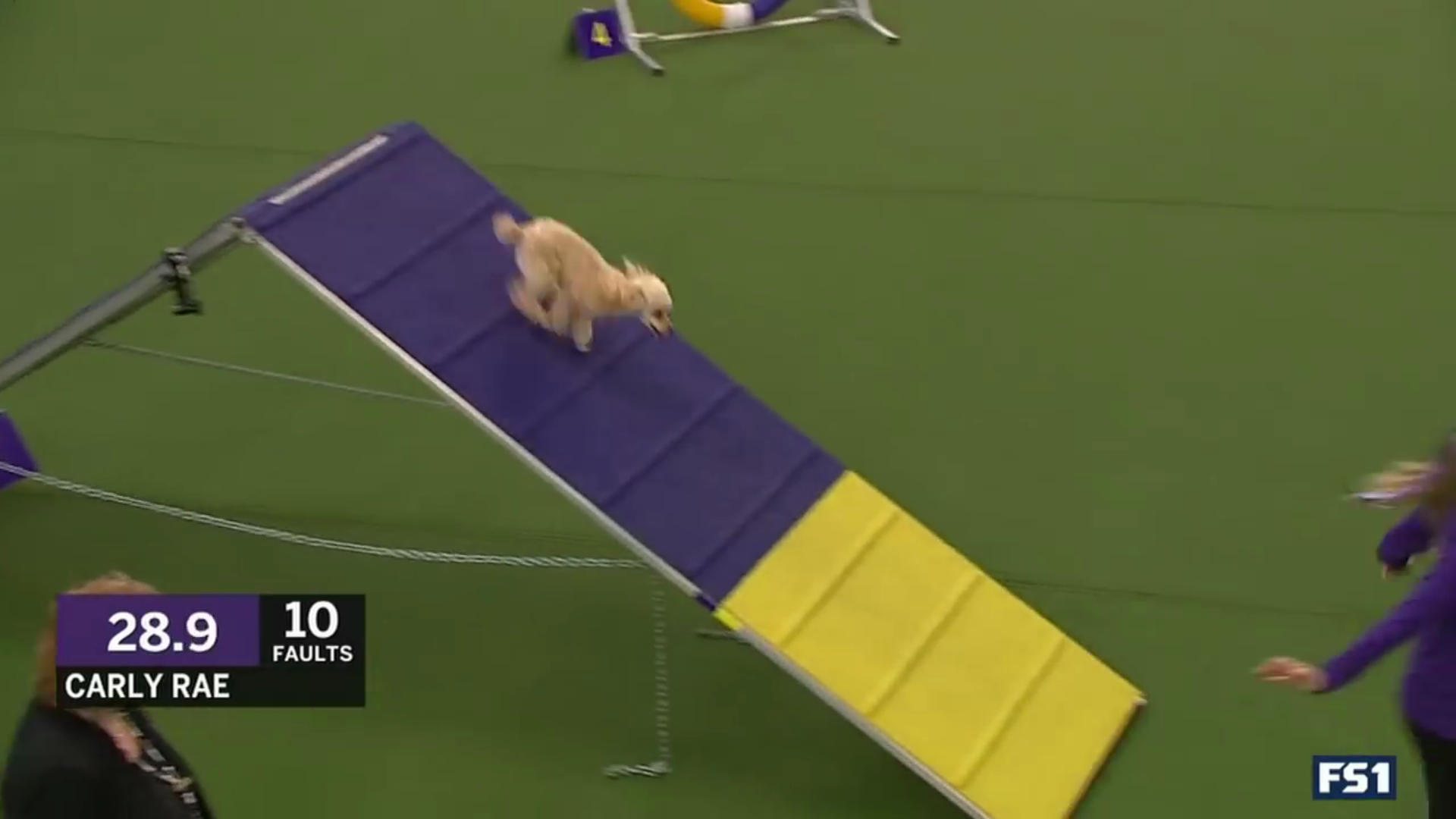}
        \label{fig:animals}
    }
    \subfigure[Health \& Fitness Video]{
        \includegraphics[width=4.0cm]{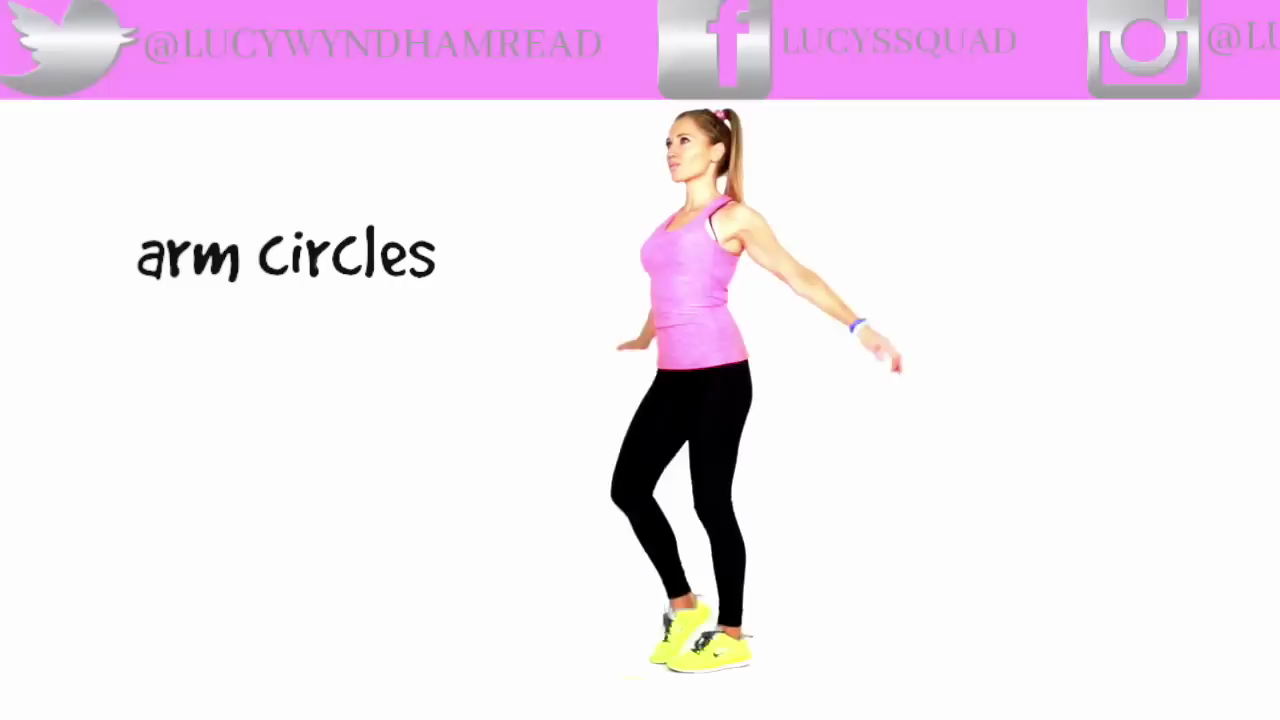}
        \label{fig:healthfitness}
    }
    % Second row of images
    \subfigure[Film and Animation Video]{
        \includegraphics[width=4.0cm]{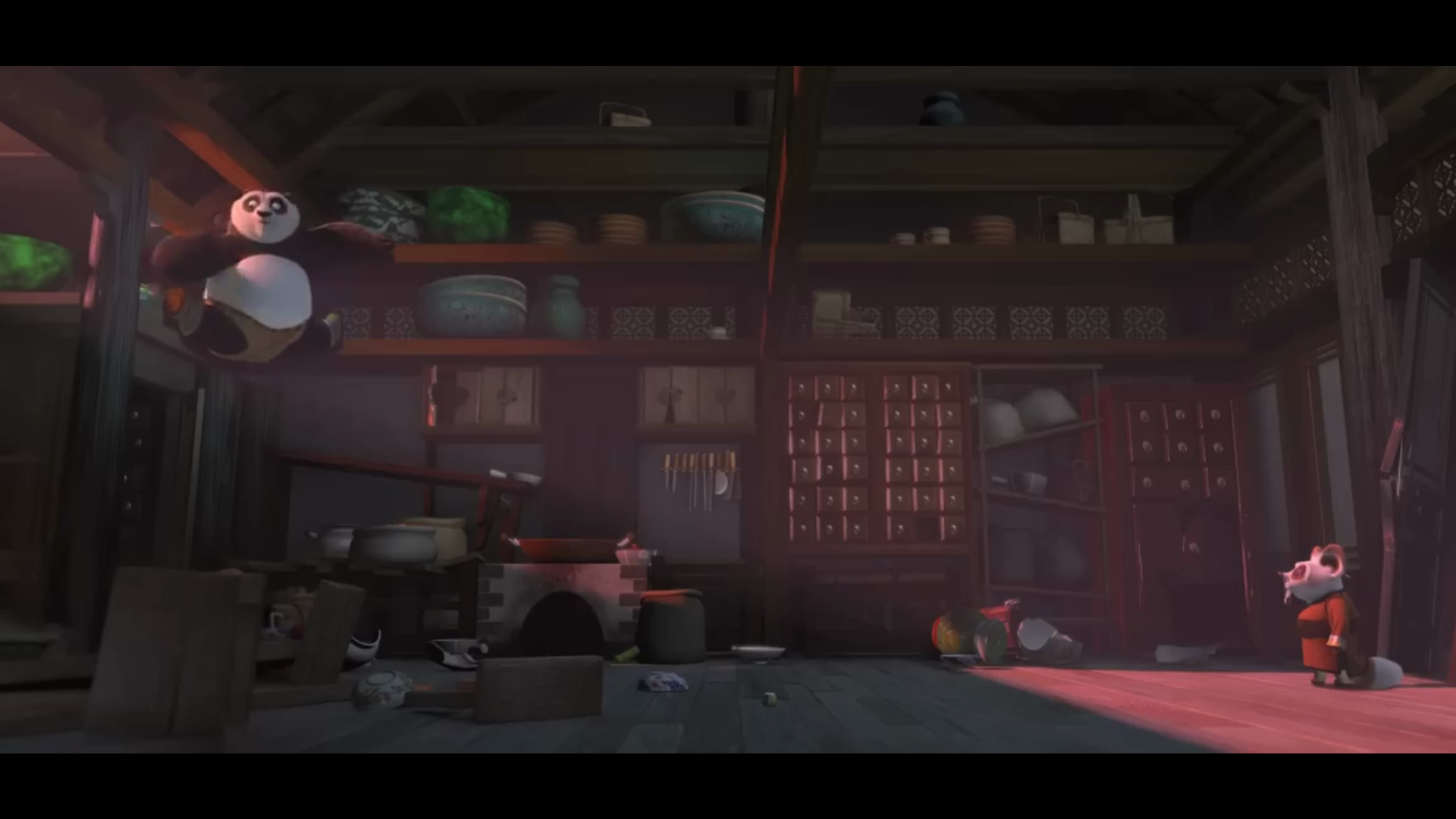}
        \label{fig:film}
    }
    \subfigure[Beauty Video]{
        \includegraphics[width=4.0cm]{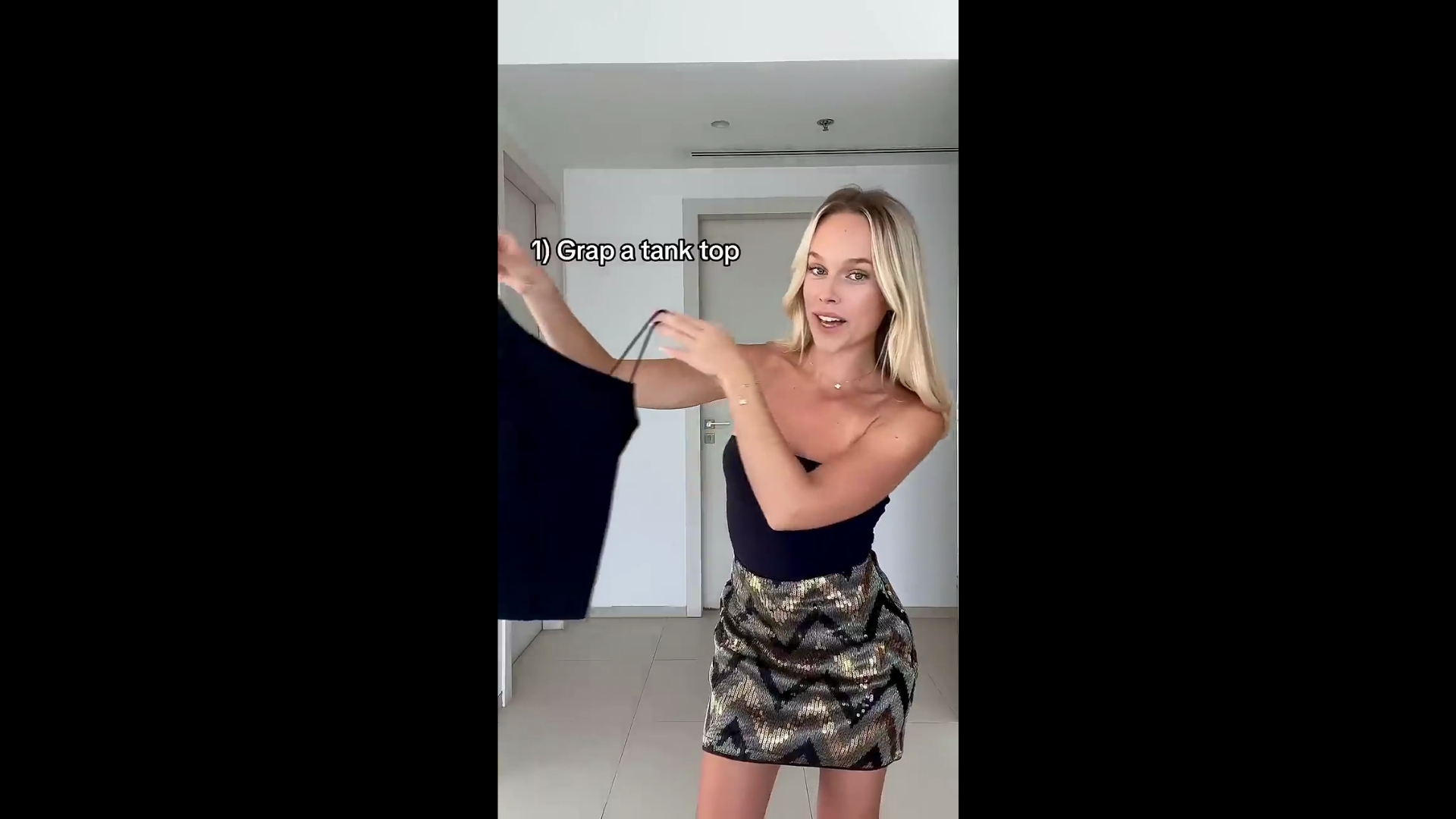}
        \label{fig:fashion}
    }
    \subfigure[Education Video]{
        \includegraphics[width=4.0cm]{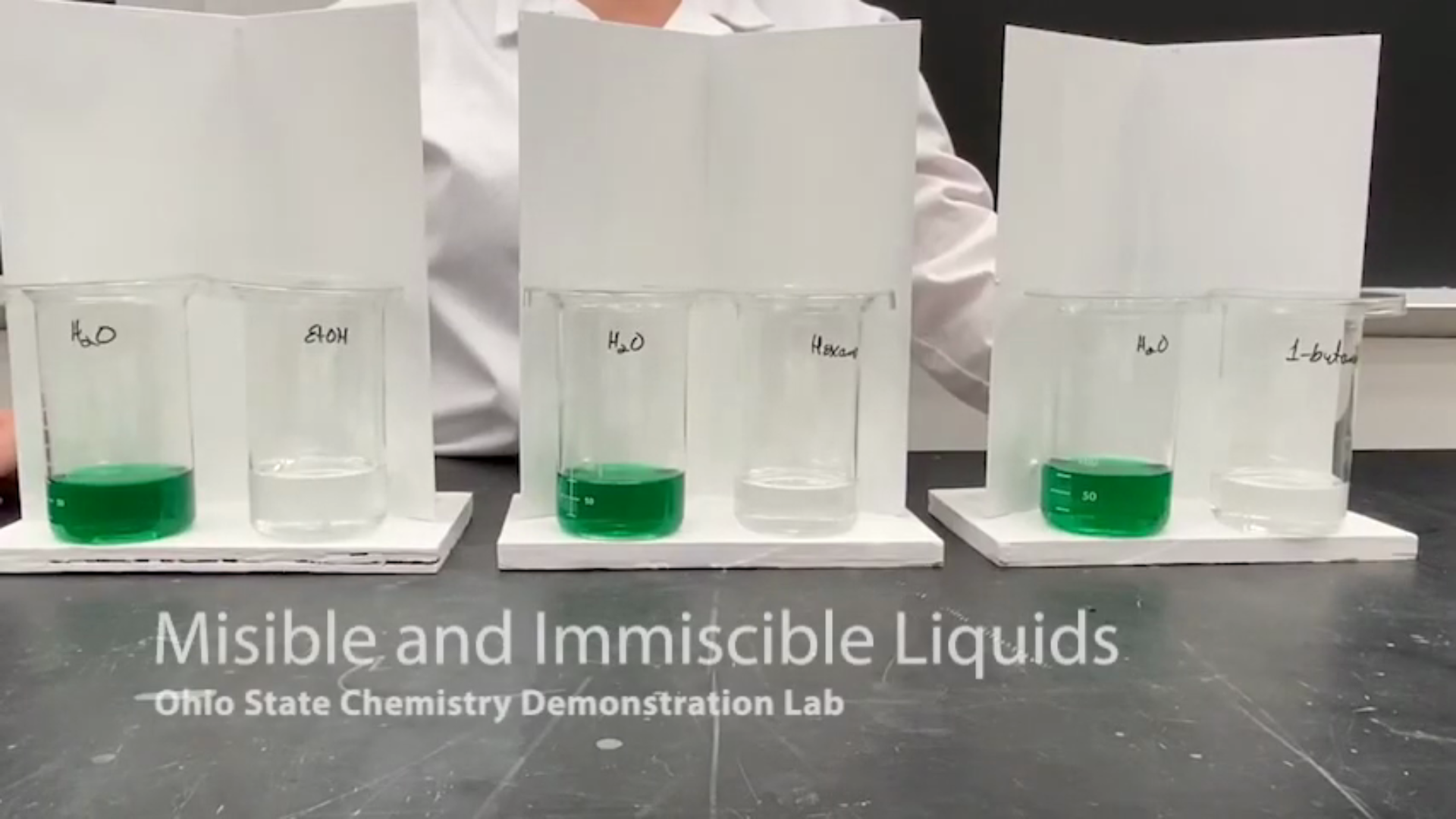}
        \label{fig:education}
    }
    \subfigure[Food \& Cooking Video]{
        \includegraphics[width=4.0cm]{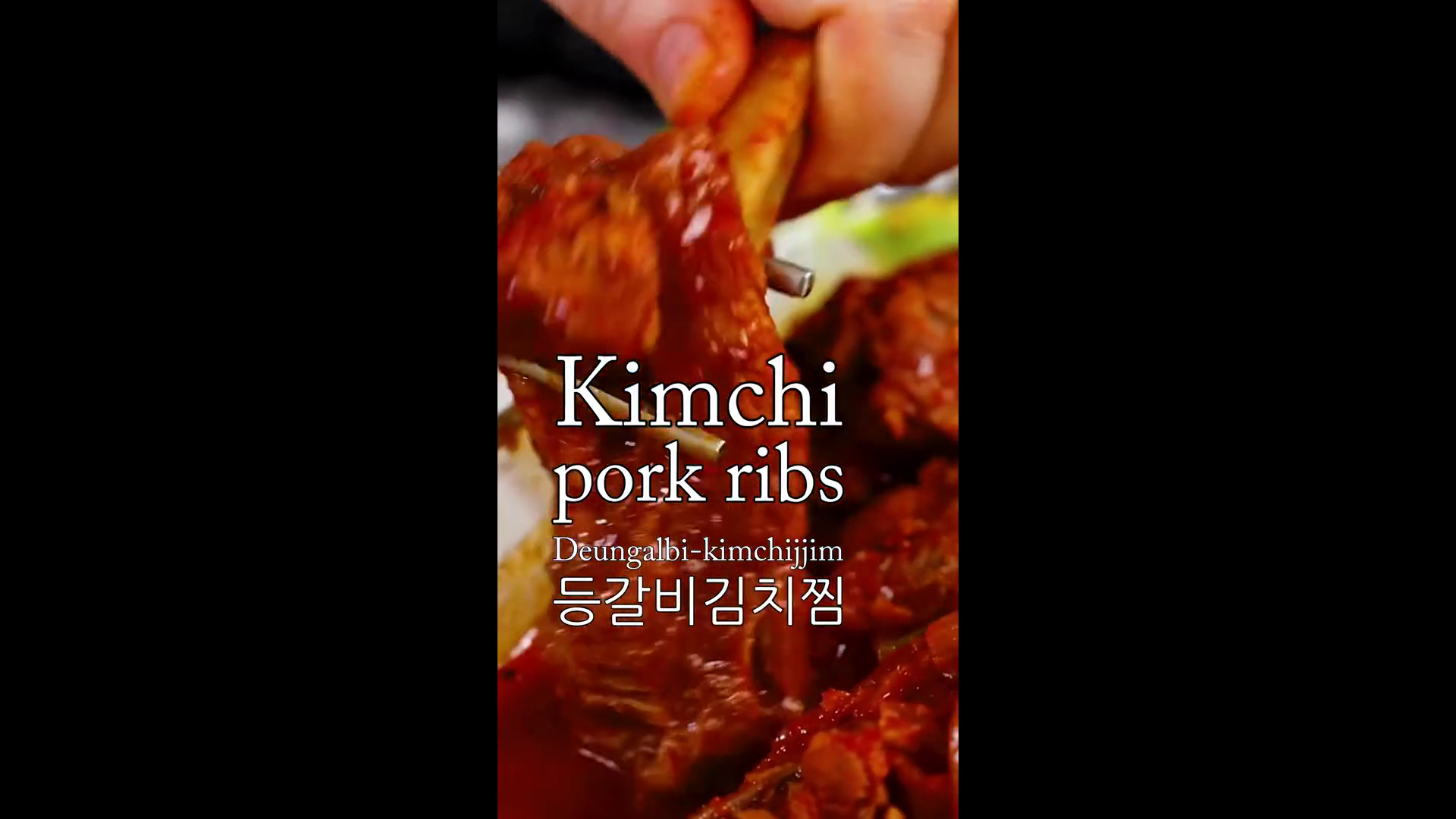}
        \label{fig:food}
    }

    \caption{Figure (a) shows the test video used to familiarize participants with activating descriptions. Figures (b) through (h) illustrate the videos included in the study. The viewing order of the seven videos was randomized for each participant in the user study.}
    \label{fig:keyframes}
    \Description{Grid of eight video thumbnails used in the study. (a) Test video showing how to make a raindrop cake. (b) People and Blogs video titled 'The Philippines in 1 Minute' featuring an aerial view of an island. (c) Pets and Animals video of a dog running on an obstacle in an agility competition. (d) Health and Fitness video showing a person performing arm circles. (e) Film and Animation video showing an animated scene from a trailer for Kung Fu Panda. (f) Beauty video of a woman in a brightly lit room discussing fashion hacks. (g) Education video demonstrating the concept of 'Miscible and Immiscible Liquids.' (h) Food and Cooking video showing a dish of kimchi pork ribs.}
\end{figure}

We chose short videos (50--110 seconds) from seven different genres to evaluate user-driven AI descriptions across a wide range of video content. The chosen genres include: Food and Cooking, Beauty, Pets and Animals, People and Blogs, Health and Fitness, Film and Animation, and Education (Figure~\ref{fig:keyframes}). We prioritized genre diversity over the number of videos per genre to account for differences in participant interests. Shorter videos were chosen to minimize the chance of disengagement due to a lack of interest in specific genres.  
%The selection of these genres was driven by the need to evaluate on-demand AI descriptions across a wide range of content. 
Our genre selection was guided by prior literature, which suggests that BLV users' needs for ADs can vary depending on the video genre~\cite{jiang2024s}. The genres also covered a range of visual elements, helping test the MLLMs' descriptive capabilities for different content. The videos in these genres varied in audio and speech content, and hence, how frequently ADs are required may vary. Additionally, videos from these genres were frequently accessed and requested on YouDescribe, making them relevant contenders for audio descriptions~\cite{you_described_we_archived}.
 
In these genres, we chose videos that contained speech and visual references and thus could benefit from ADs. We did not include videos that relied primarily on speech or had a music soundtrack. Our initial pilot study with a blind user on 14 different videos suggested that the former category was already accessible and did not require ADs, and the latter category was difficult to use with user-driven ADs due to the limited auditory cues. Thus, we selected videos with a speech track where visual elements were important (e.g., a workout video) or were referenced without explanation~\cite{video_inaccessible}. For example, in the \emph{Beauty} video, the speaker said, ``stop wearing your top like this,'' while pointing to her top without describing its appearance (Figure~\ref{fig:teaser}). 

\subsection{Audio Description Generation}
Our goal was to evaluate the effectiveness of user-driven AI descriptions as a tool for BLV users to watch and interact with videos. However, generating descriptions and speech from visual frames can introduce delays and negatively impact the user experience. Moreover, the output of MLLMs is not deterministic and can change on each API call. Thus, we pre-generated ADs for the videos to run a controlled experiment and minimize the influence of these factors on the results. To generate the descriptions, we collected AD guidelines from online sources and prompted GPT-4V with the guidelines and video frames.
% and post-processed the descriptions for consistency.

\textbf{Collecting AD guidelines.} We collected AD guidelines from four online sources. %These guidelines are targeted towards describers on what are some of the best practices for creating audio descriptions. The research team gathered guidelines from different online sources. 
These guidelines focused on how to describe educational content ~\cite{dcmp_description_key}, entertainment content~\cite{mediaccess,netflix_style_guide}, and general guidelines for audio describers~\cite{ofcom_access_services}. An example guideline was \textit{``Description should convey facial expressions, body language, and reactions, especially when in opposition to the dialogue. These elements can be omitted if they completely mimic the dialogue they are accompanying.''}~\cite{netflix_style_guide}. We collected an initial list of 154 AD guidelines. Repeated guidelines and those
%the ones that GPT-4V could not follow such as \emph{``Description should include known relationships when they have been revealed.''} (Netflix). All guidelines 
that focused on timing, context, and audio content of the video were removed, as these aspects cannot currently be used to prompt GPT-4V. %We also shortened the guidelines for brevity. 
% For example, \textit{``Avoid over-describing — do not include visual images that are not vital to the understanding or enjoyment of the scene''} was shortened to \textit{``Avoid over-describing — do not include non-essential visual details.''}. 
This process resulted in 42 guidelines for prompting MLLMs (Appendix~\ref{app:ad}). The guidelines were then passed along with the prompt (Appendix~\ref{app:prompt}) and video frames to create descriptions. %Some of the guidelines were shortened. 

%Once collected, the guidelines were categorized using a codebook, one of the codes in the codebook was \textbf{\textit{``GPT prompt''}}, which was used to categorize whether GPT could adhere to the guideline or not to create description. This removed guidelines such as ``Description should include known relationships when they have been revealed.'' (Netflix). All guidelines that focused on timings, context and audio content of the video were removed as they currently cannot be used to prompt models. This left a total of 42 guidelines. Some of the guidelines were shortened. For example, \textit{“Avoid over-describing — do not include visual images that are not vital to the understanding or enjoyment of the scene}” was shortened to \textit{``Avoid over-describing — do not include non-essential visual details.''}

\textbf{Generating descriptions.} For each video, we generated two types of ADs, a shorter version (\emph{concise}) and a longer version (\emph{detailed}), %Jiang et al. (2024) have emphasized the importance of allowing BLV users to choose the level of detail as preferences with BLV users can vary. Additionally, user goals can vary based on the type of video. For instance, a cooking video only requires descriptions related to following the recipe, meanwhile for a travel blog, BLV users might require more detail on the visuals for a more immersive experience. To cater to this, we wanted the BLV participants to be able to get either detailed or concise on demand descriptions based on their preference.
to let BLV participants choose the level of detail based on their preferences and video content.
In our experimentation, directly prompting GPT-4V to create concise and detailed descriptions led to inconsistent results on different API calls. Thus, we created two versions by specifying the maximum length of descriptions in the input prompt: 100 words for detailed descriptions and 25 words for concise descriptions. To determine the length, we evaluated maximum description lengths of 25, 50, 75, 100, and 200 words. The upper limit of 200 sometimes led to excessively verbose and repetitive descriptions. Also, when the lengths of concise and detailed were close (e.g., 50 and 75), the descriptions were often hard to differentiate. Based on iterative prompting and testing, we found 25 and 100 words
%Although the descriptions were much shorter than the word limit, based on testing this was
the most effective in creating the two types of descriptions without altering the prompt. 
% An example of concise vs detailed for a frame are visible in Figure~\ref{fig:ADexample}.
For a user-driven experience, we generated a description for every second of the video using GPT-4V API. Each API call consisted of the prompt and ten frames from the video, with one frame extracted from each second. Thus, each API call returned ten descriptions that describe the ten consecutive seconds of the video, requiring six API calls (i.e., 60 descriptions) to describe a one-minute video. %Ten frames were sent in each API call so that the LLM had context about the video. 
We repeated the process twice for the seven videos to get detailed and concise descriptions. The same AD guidelines were used for both types of descriptions. %However, the detailed ADs contained more visual detail and described the on-screen text which was often omitted from concise ADs (Appendix~\ref{app:descriptions}).
% During the AD generation process, there were a few instances when the generated descriptions referenced previous descriptions. For instance, in the \emph{Film and Animation} video, the description for a frame started as: \emph{``Same as description 5''}. Because these errors were few and required manual correction for removal, we decided to keep them in the descriptions and assess their impact on BLV users' experience.

\textbf{Reviewing description quality.} We checked all the descriptions to assess the concise and detailed descriptions for the same scenes and check the overall description quality. The detailed ADs contained more visual detail about the scene, object properties, and people's appearances and described the on-screen text which were often omitted from concise ADs (see Appendix~\ref{app:descriptions}). In AD generation, there were a few instances in which the generated descriptions referenced previous descriptions. For instance, in the \emph{Film and Animation} video, the description of a frame started as: \emph{``Same as description 5''}. Additionally, because each API call generated descriptions independently without context from previous API calls for the same video, this sometimes resulted in the repetition of descriptions of the same characters and visual detail. The lack of context also resulted in hallucinations. For example, in the cooking video, the pork ribs were misidentified as chicken wings in the latter half of the video. Because these errors were few and required manual correction for removal, we decided to keep them in the descriptions and assess their impact on BLV users' experience.

\subsection{User Interface}
\begin{figure*}[htbp]
    \centering
    \includegraphics[width=0.7\textwidth]{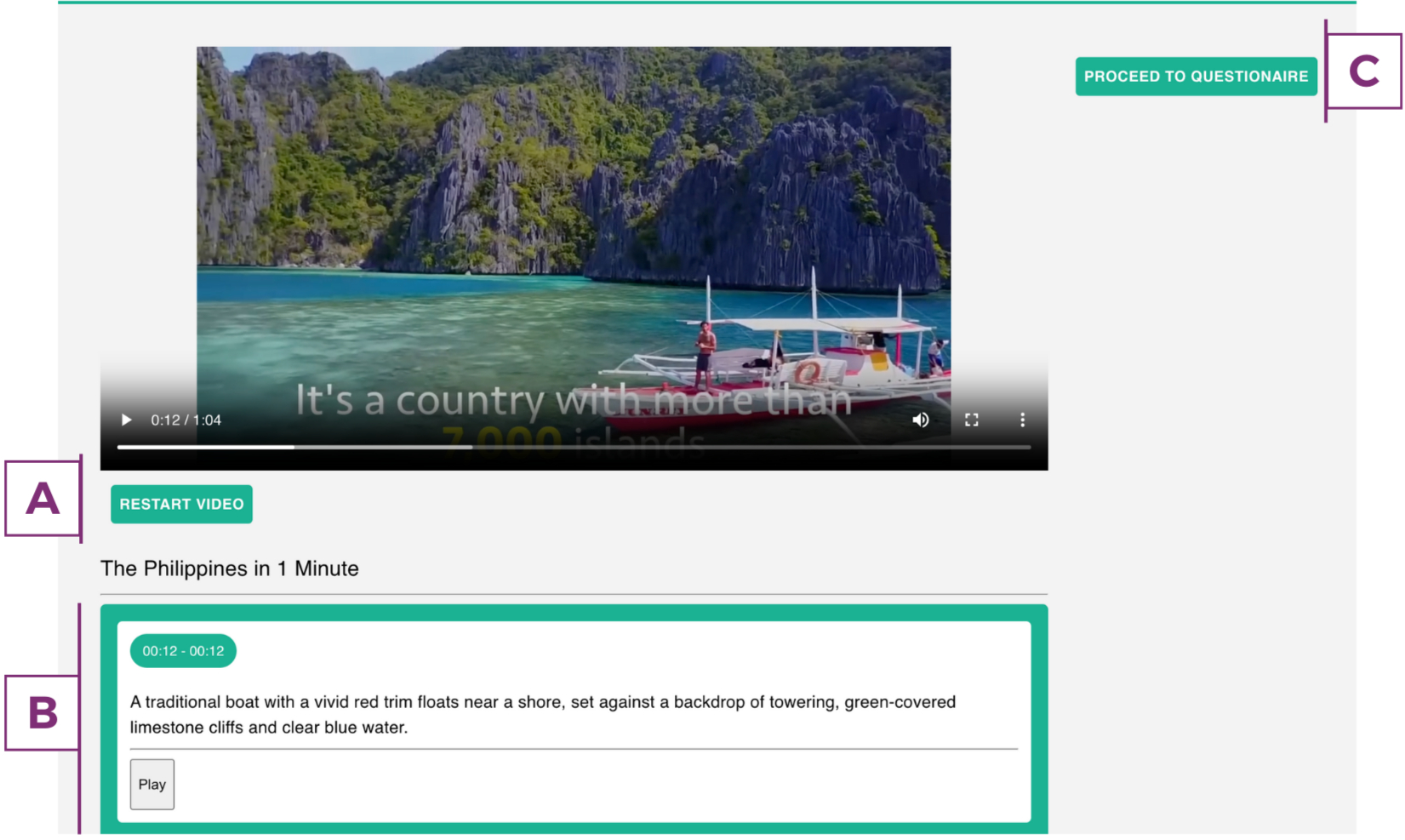}
    \caption{Interface to test user-driven descriptions with (A) a video viewing section and video restart button, (B) a description box displaying the current timestamp, description, and a play button to replay the description, and (C) a button to proceed to the questionnaire and rate their experience of user-driven interaction with the video. When the user activates the description by pressing ``C'' for concise or ``D'' for detailed on the keypad, the video pauses to play the description and resumes once the description is over. } 
    \label{fig:interface}
    \Description{User interface for the prototype used to test user-driven descriptions. The interface displays the People \& Blogs video titled 'The Philippines in 1 minute,’ which shows a scene of a beach and mountains with people on boats. Below the video is a description box with text corresponding to the description activated by a user. Buttons include 'RESTART VIDEO', 'Play', and 'PROCEED TO QUESTIONNAIRE.'}
\end{figure*}

We prototyped a simple interface to test the user-driven AI descriptions (Figure~\ref{fig:interface}). We used ReactJs for creating the interface and Firebase for storing all the generated ADs and videos. %To reduce time delays, all the ADs were generated beforehand and stored in the Firebase database. 
Participants could press ``C'' on their keyboard to get concise descriptions and ``D'' for detailed descriptions. We used an extended audio description approach, where the video paused when the participant pressed a key, the text-to-speech model read out the description, and then the video playback continued. We used OpenAI's \texttt{alloy} text-to-speech model to read out the descriptions with a natural human-like tone. The interface logged the key presses (C or D) and their timings. After watching a video, participants could rate their experience and type questions on the interface. 

\section{User Study}%USER STUDY}
We conducted a user study to %examine how BLV users use and perceive on-demand ADs for various videos and  %how well the on-demand AI descriptions performed based on the metrics of enjoyable, efficiency and effectiveness in understanding visual content. 
elicit user perception of this interactive approach to watching videos and obtain data on the frequency and type of AD requests across different videos and BLV individuals. 
The study was approved by our institution's IRB. The study was conducted over Zoom to facilitate participation from different locations. Each study session took about 75 minutes, and participants received \$50 for their time.

\begin{table*}
\footnotesize
  \caption{Description of the 20 BLV participants in our user study. The participant numbers are sub-indexed to indicate their vision status: Blind (B), Legally Blind (LB), and Low Vision (LV)}
  \label{tab:Demographic Information}
  \centering
  % \resizebox{\textwidth}{!}{%
  \begin{tabular}{p{0.5cm}p{0.3cm}p{1.5cm}p{0.5cm}p{7.5cm}p{3.5cm}p{1cm}}
 % {ccccccc}
    \toprule
    \textbf{P\#} & \textbf{Age} & \textbf{Race} & \textbf{Gender} & \textbf{Visual Impairment} & \textbf{Screen Readers} & \textbf{AD Use} \\
    \midrule
    P1$_B$  & 63 & White & Female & Total vision loss due to car accident & VoiceOver, JAWS & Frequently \\
    P2$_B$  & 57 & Unknown & Male & Blind due to retina detachment and glaucoma %most of life 
    & JAWS & Frequently \\
    P3$_{LV}$  & 68 & Black & Female & Low vision since birth & VoiceOver, Zoom, Magnifying glass & Occasionally \\
    P4$_B$  & 32 & Unknown & Female & Totally blind due to retinoblastoma as an infant & JAWS & Occasionally \\
    P5$_B$  & 64 & White & Male & Totally blind, born with glaucoma, had some sight until the age of 15. Lost sight at 15 and have no light perception & VoiceOver, JAWS & Frequently \\
    P6$_B$  & 44 & White & Male & Totally blind & VoiceOver, JAWS & Frequently \\
    P7$_B$  & 28 & Asian & Male & Total congenital blindness & JAWS & Occasionally \\
    P8$_B$  & 39 & White & Male & Totally blind from retinal arterial occlusion & VoiceOver, JAWS & Frequently \\
    P9$_B$ & 46 & White & Male & Congenital blindness due to glaucoma and optic nerve damage & VoiceOver, JAWS & - \\
    P10$_{LV}$ & 79 & White & Male & Low vision & VoiceOver, Mechanical magnifier & Occasionally \\
    P11$_{LV}$ & 27 & Hispanic/Latino & Male & Low vision since birth & VoiceOver, JAWS, Talk Back & - \\
    P12$_{LB}$ & 40 & American Indian & Female & Slowly losing vision due to stargardt disease & JAWS & - \\
    P13$_B$ & 30 & White & Female & Totally blind & JAWS, NVDA & Rarely \\
    P14$_{B}$ & 59 & Black & Male & Retinitis Pigmentosa, birth defect that led to progressive visual loss with age & JAWS, NVDA & Frequently \\
    P15$_{LB}$ & 27 & White & Male & Legally blind since birth, no vision in left eye & VoiceOver, Screen Magnification & Frequently \\
    P16$_B$ & 38 & White & Female & Total congenital blindness & JAWS, NVDA & Frequently \\
    P17$_{LB}$ & 44 & Hispanic/Latino & Male & Legally blind, Light perception only %due to degenerative eye disease, 
    Retinitis Pigmentosa & JAWS, NVDA & Frequently \\
    P18$_B$ & 30 & Asian & Male & Totally blind, almost since birth & VoiceOver, JAWS & Frequently \\
    P19$_B$ & 42 & Asian & Male & Totally blind caused by scarlet fever & VoiceOver, NVDA & Frequently \\
    P20$_B$ & 32 & Hispanic/Latino & Female & Total congenital blindness & VoiceOver, NVDA, JAWS & Frequently \\
    \bottomrule
  \end{tabular}%
  % }
  \label{tab:participants}
\end{table*}

{\bfseries Participants.} We recruited 20 participants (7 female, 13 male) through BLV organizations, Facebook groups, and snowball sampling. The participants were between 27 to 79 years old and had various vision impairments ranging from legal blindness, blindness with some light and color perception, and total blindness (Table~\ref{tab:participants}). Most participants ($n=18$) used screen readers to navigate the interface, while two participants used magnification. %Table~\ref{tab:participants} presents the demographics of all participants.

\begin{figure*}[htbp] 
    \centering
    \includegraphics[width=0.95\textwidth]{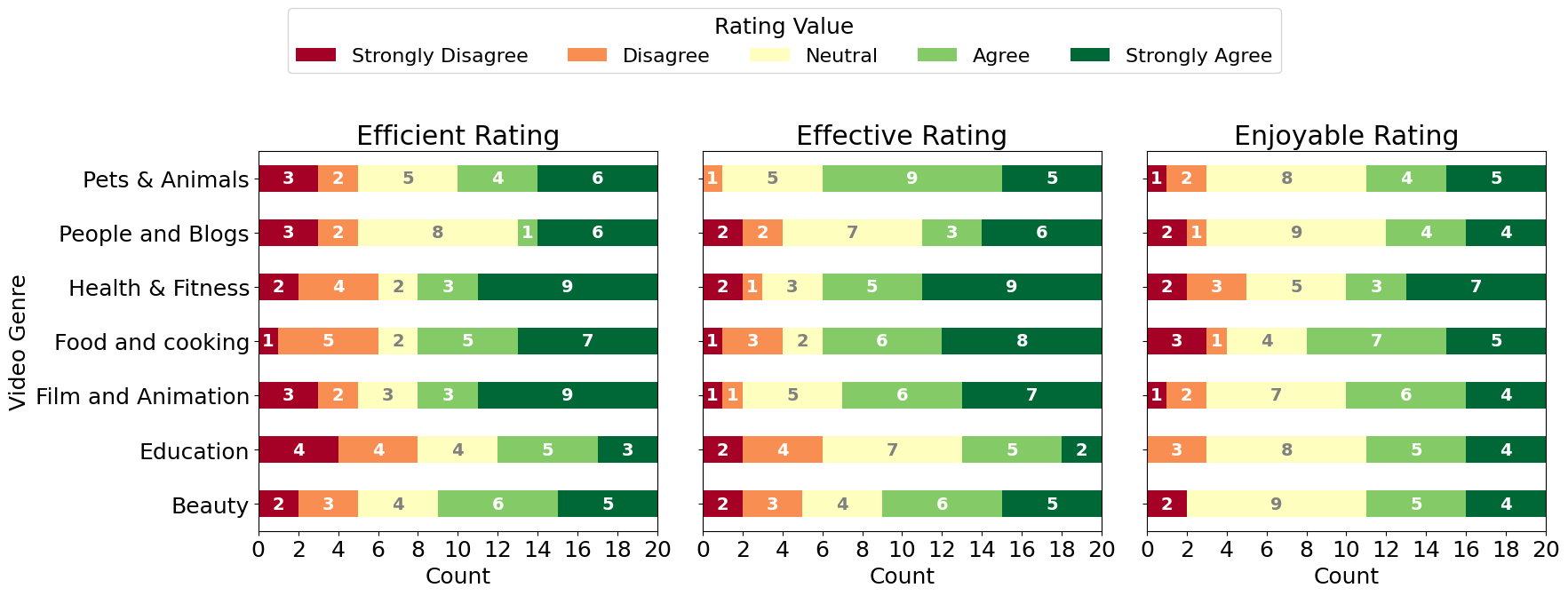}
    \caption{Distribution of Participant Ratings Across Video Genres}
    \label{fig:ratings}
    \Description{Three bar charts showing the distribution of participant ratings across different video genres for three ratings: Efficient Rating, Effective Rating, and Enjoyable Rating. The genres include Pets & Animals, People and Blogs, Health & Fitness, Food and Cooking, Film and Animation, Education, and Beauty. Rating values segment each bar: Strongly Disagree, Disagree, Neutral, Agree, and Strongly Agree.}
\end{figure*}

{\bfseries Procedure.}
The study session included four phases: completing a demographics survey, a practice segment, a video viewing segment with user ratings, and a semi-structured interview to gather qualitative feedback.
%We used a mixed design approach which consisted of two segments: a video viewing segment with user ratings and an interview segment.  %The study was conducted over zoom and took approximately 75 minutes. 
After completing the demographic survey, participants interacted with a testing page to get familiar with the interface and used C or D on the keypad to activate descriptions. %The video on the testing page was a one-minute cooking video that contained no speech to encourage the AD requests. Once the participants were comfortable with the user-driven interaction, they could proceed to the user study. 
Then, participants watched the seven videos in random order with user-driven descriptions. We instructed the participants to watch the videos to understand the content so they could summarize the video. They watched each video once and pressed the C and D keys whenever needed to receive an AD for the previous second in the video. The UI logged the time and type of key presses.
To evaluate user-driven descriptions, after each video, participants rated their experience on a 5-point Likert-type scale from ``strongly disagree'' to ``strongly agree''. The ratings assessed (a) {Efficiency}: ``The audio descriptions made watching the video more efficient,'' (b) {Effectiveness}: ``The information provided in the audio descriptions was effective in helping to understand the visual content of the video,'' and (c) {Enjoyability}: ``The experience of listening to the audio descriptions while watching the videos was enjoyable''. 
% The ratings assessed the efficiency of watching the video with AD, the enjoyability of listening to AD while watching the video, and the effectiveness of user-driven ADs in understanding the visual content. 
For three of the seven videos, participants were asked to provide a brief summary of the content to encourage users to carefully watch and engage with the videos. Participants could also type any questions related to the video. Finally, we conducted a semi-structured interview with participants to gather in-depth feedback on their experience with user-driven AI descriptions. %(see Appendix~\ref{app:interview} for interview questions). 
The interviews were audio-recorded and transcribed.

{\bfseries Data Analysis.} We analyzed the interview transcripts using thematic analysis inspired by Braun and Clarke's approach~\cite{clarke2021thematic}. Specifically, two authors independently applied open coding to each transcript using MAXQDA qualitative analysis software. After coding every five interviews, the two authors met and discussed the similarities and differences between their codes before moving on to the rest. The authors also wrote memos to capture interesting patterns and relationships between codes. After open-coding all the transcripts, the two authors separately identified recurring patterns or themes from the codes and discussed the themes together. One author wrote a draft of the themes, and both authors discussed and revised it by referencing the codes. This process resulted in three themes, each including 2-4 sub-themes (Section~\ref{sec:themes}). 
% We also applied statistical testing on the user ratings and provide summary statistics for timing and type of AD requests (Section~\ref{sec:quant}). %themes were discussed and iterated by the two authors. We highlight the themes from the analysis in the quantitative section.

\section{Results} %RESULTS AND FINDINGS}
This section presents quantitative results followed by qualitative themes from the interviews.

\subsection{Quantitative Results}
\label{sec:quant}

\subsubsection{\textbf{User Ratings}} %On average, the participants rated the efficiency of on demand way of receiving descriptions at 3.42 on a scale from 1 (strongly disagree) to 5 (strongly agree). %The effectiveness of the descriptions in helping them understand the visual content was rated slightly higher at 3.64. Meanwhile the enjoyability of the experience was rated at 3.47. 
Figure~\ref{fig:ratings} shows the distribution of user ratings for the efficiency in watching, effectiveness in comprehending the video, and enjoyability of videos with the user-driven ADs. The medians for efficiency and effectiveness are 4 (agree), while the median for enjoyability is 3 (neutral). The lower median for enjoyability compared to efficiency and effectiveness suggests that while the majority agreed that descriptions were useful, there was less consensus on overall enjoyment and many found the experience emotionally neutral. 
Participants noted that the extended (rather than inline) presentation of user-driven ADs reduced their enjoyment of the videos. Statistical testing with the Friedman test did not show a significant effect of genre on user ratings.

\subsubsection{\textbf{Frequency and Type of AD Requests}}

% We conducted a repeated measures ANOVA to determine statistical differences in the time intervals between subsequent AD activations (Figure~\ref{fig:interval}). %We can report that, with sphericity assumed, 
% The mean scores of the AD intervals were statistically significantly different, $F(6, 114) = 5.460$, $p < .001$, $\eta^2$=0.223. Post hoc pairwise comparison using the Bonferroni correction showed a significant difference in the AD intervals for \emph{Education} vs. \emph{Film and Animation} ($p = 0.012$), \emph{Film and Animation} vs. \emph{Health and Fitness} ($p = 0.009$), \emph{Beauty} vs. \emph{Film and Animation} ($p < 0.001$), and \emph{Education} vs. \emph{Pets and Animals} ($p = 0.046$). The results suggest that the frequency of descriptions required can vary for different video genres. 

\begin{figure*}[htbp]
    \centering
    \subfigure[The time intervals between AD activations]{
        \includegraphics[width=0.48\textwidth]{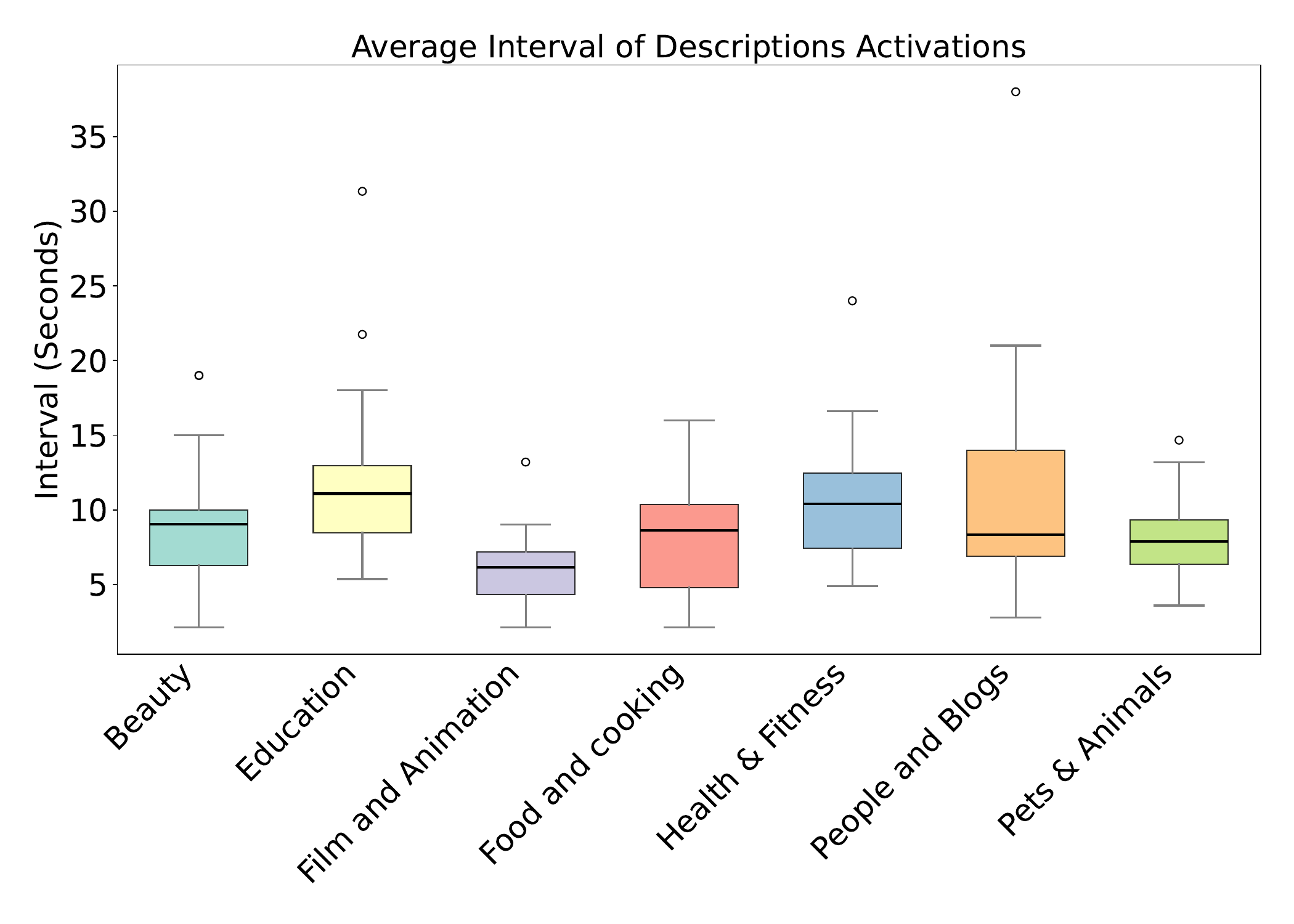}
        \label{fig:interval}
    }
    % \hfill
    \subfigure[Number of concise and detailed AD activations]{
        \includegraphics[width=0.48\textwidth]{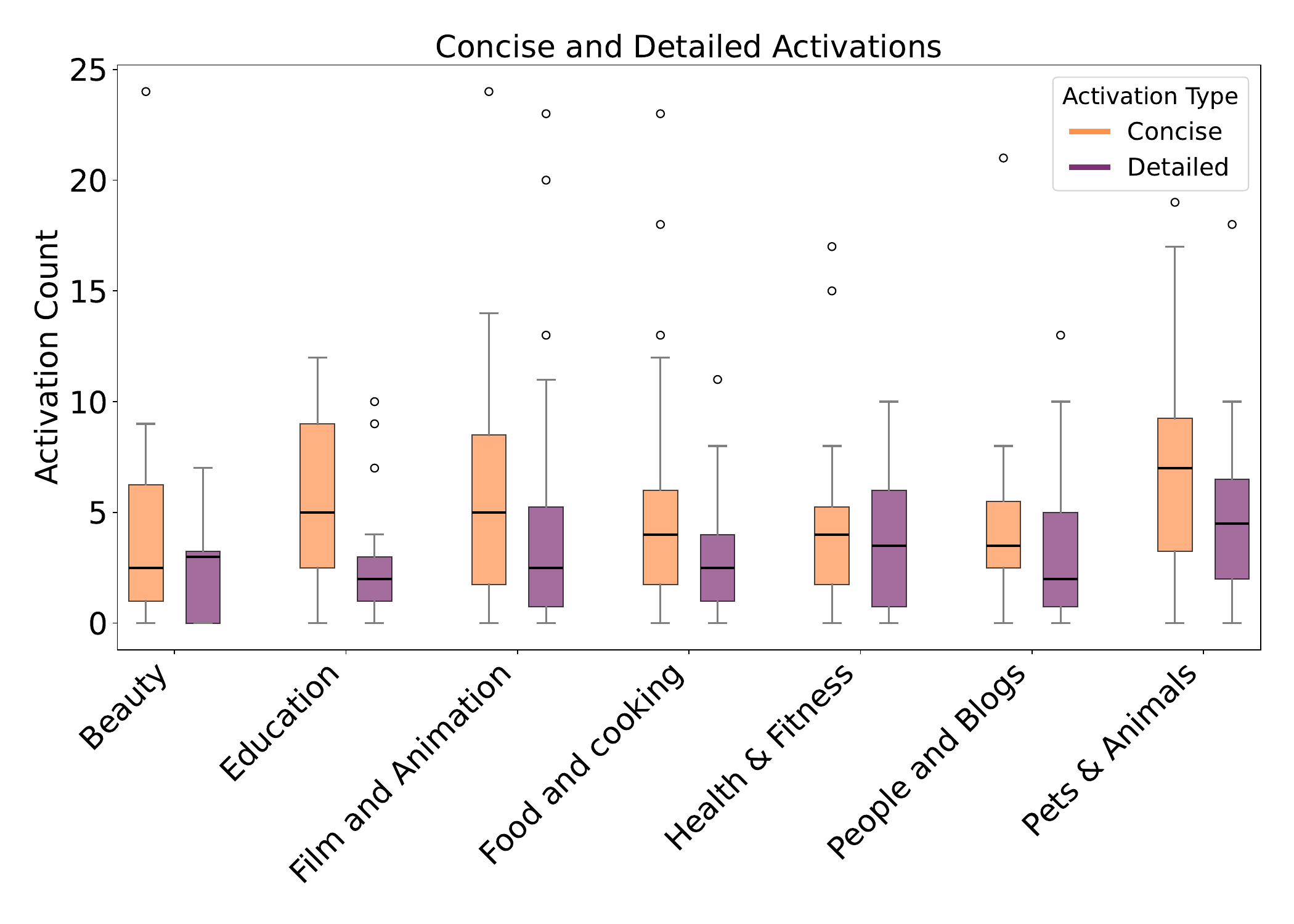}
        \label{fig:requests}
    }
    \caption{Results for the frequency and type of AD activations in the seven videos. }
    \label{fig:comparison}
\Description{Two box plots showing results for the frequency and type of audio description (AD) activations across seven video genres. Plot (a) illustrates the average time intervals between AD activations for the seven video genres. %Significant differences are indicated by asterisks for the following: Film and Animation vs. People and Blogs, Film and Animation vs. Health & Fitness, Beauty vs. People and Blogs, and Education vs. Health & Fitness, with more asterisks indicating higher significance. 
Plot (b) shows the number of concise and detailed AD activations, represented by orange and purple boxes, respectively.}
\end{figure*}

Figure~\ref{fig:interval} shows the average time interval between activations for different genres. On average, activation intervals varied across genres, with more frequent activations for \emph{Film and Animation} ($Mean = 5.9s$, $SD=2.6$) and \emph{Pets \& Animals} ($Mean = 7.9s$, $SD=2.9$), and less frequent activations for genres like \emph{Education} ($ Mean = 12.3s$, $SD=5.9$) and \emph{Health and Fitness} ($ Mean = 10.5s$, $SD=4.6$). These patterns suggest that some video genres may require more frequent descriptions based on the audio and visual content. For the number of AD activations, overall concise descriptions were activated more frequently ($Mean=5.42$ , $SD=5.26$) than detailed ones ($Mean=3.58$ , $SD=3.95$) in all videos (Figure~\ref{fig:requests}). We focus on providing summary statistics instead of running statistical tests since these quantitative patterns may not generalize to a larger sample size of BLV users and variations in video genres. %For the number of AD activations, the assumption of normality was violated. Thus, we conducted Aligned Rank Transform (ART~\cite{ARTool}), a non-parametric alternative to two-way repeated measures ANOVA, to investigate the effects of request type (concise, detailed) and video genre on the counts of AD activations (Figure~\ref{fig:requests}). The description type had a significant main effect $F(1, 247) = 12.52$, $p = 0.00048$, indicating concise descriptions were activated more frequently ($Mean=5.42$ , $SD=5.26$) than detailed ones ($Mean=3.58$ , $SD=3.95$) for all videos. 

% There was no significant effect of video genre $F(6, 247) = 2.11$, $p = 0.053$, and no interaction effect between video genre and the type of requests $F(6, 247) = 0.44$, $p = 0.85$.

% To assess the effect of video genre on the type of requests, we conducted two repeated measures ANOVAs to on the number of requests for concise and detailed descriptions with one factor of video genres. With Greenhouse-Geisser correction, there were significant differences for concise requests $F(3.272, 62.163) = 2.807$, $p = .042$, $\eta_p^2$=0.129, and detailed requests $F(3.267, 62.069) = 3.110$, $p = .029$, $\eta_p^2$=0.141, based on the video genre. Despite an overall significant effect, pairwise comparisons did not reveal statistically significant differences between genres for either concise or detailed requests.

% To examine the effects of requests type and video genre, we conducted repeated measures ANOVA on number of requests with two factors, video genre and type of description request. The interaction between request type and video genre was not statistically significant, $F(3.186, 60.542) = .606$, $p = .624$, $\eta^2$=0.031 (Greenhouse-Geisser corrected). The results indicate that the effect of request type on the number of requests across video genres did not vary significantly.

\begin{figure*}[h] %htbp]
\includegraphics[width=\textwidth,height=1.7in]{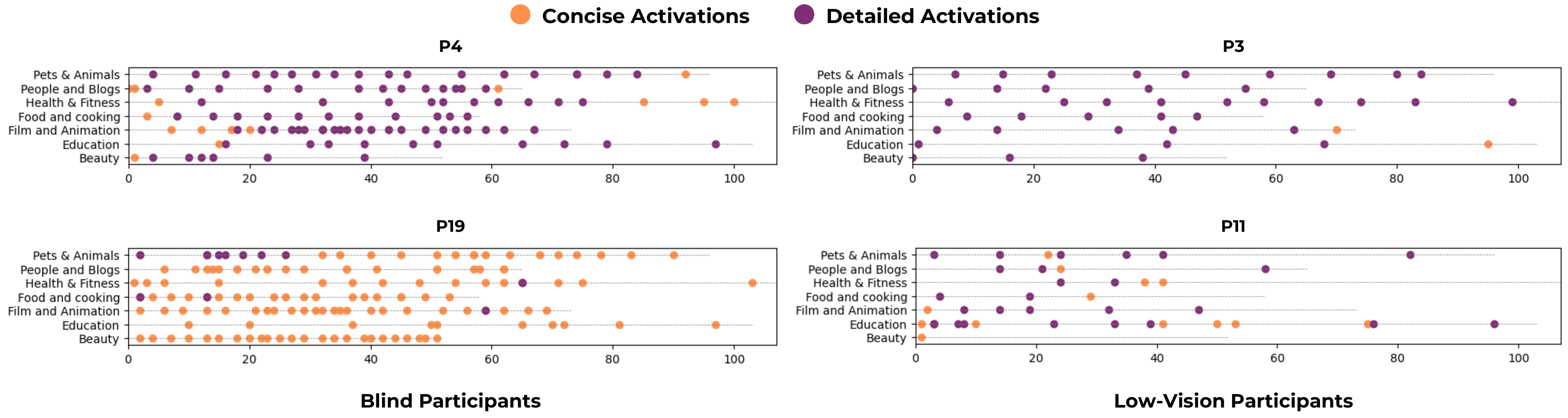}   \caption{Patterns of concise and detailed AD activations from example blind and low vision participants. On each plot, the rows show the seven videos, and the horizontal axis shows the video timeline in seconds. The video durations were between 52 and 107 seconds.} 
  \Description{Plots showing the patterns of concise and detailed description activations from four participants, split between blind and low vision. The participants include P4, P19 (blind participants) and P3, P11 (low vision participants). Each plot shows the timeline of seven videos (Pets \& Animals, People and Blogs, Health \& Fitness, Food and Cooking, Film and Animation, Education, Beauty) along the horizontal axis, with orange dots representing concise activations and purple dots representing detailed activations. The x-axis represents the video timeline in seconds. The duration of the videos varied between 52 and 107 seconds.}
~\label{fig:examplerequests}
\end{figure*}

The patterns of AD activations showed variations among BLV individuals (Figure~\ref{fig:examplerequests}).
Overall, the frequency of AD activations depended on the video type and the participants' interests. 
For example, P4$_B$ had fewer AD activations for the \emph{Beauty} and \emph{Education} video. 
The frequency and type of AD requests also varied drastically depending on the BLV individuals.
Some participants activated ADs frequently (e.g., P4$_{B}$ and P19$_{B}$), whereas others activated AD less often (e.g., P3$_{LV}$, P11$_{LV}$). 
Some participants primarily used either concise or detailed ADs (e.g., P3$_{LV}$, P4$_{B}$, P19$_{B}$), whereas some others used both types of ADs depending on the video content (e.g., P11$_{LV}$). Furthermore, the low vision participants in our study activated ADs less than the blind participants, perhaps due to relying on their functional vision. These variations further suggest the differing AD needs and preferences of BLV individuals.

\subsection{Qualitative Findings}
\label{sec:themes}

\begin{figure*}
\includegraphics[width=\textwidth]{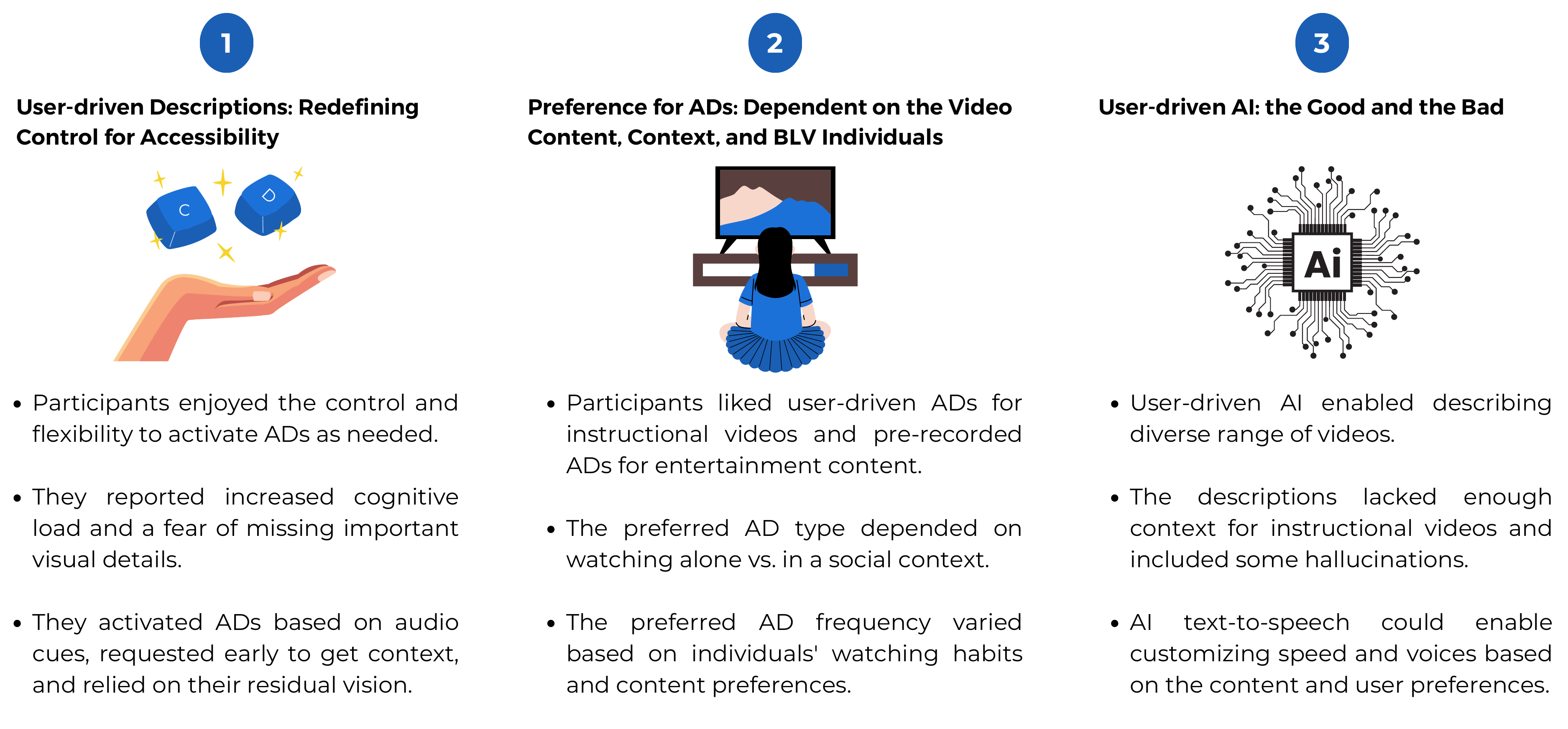}   
\caption{Overview of the three themes and sub-themes from interviews about BLV users' experience with user-driven AI descriptions.} 
\Description{Infographic summarizing the three key themes from interviews with blind and low vision (BLV) users' experiences with user-driven AI descriptions. The first theme, 'User-driven Descriptions: Redefining Control for Accessibility,' highlights participants' enjoyment of control and flexibility in activating descriptions and challenges like cognitive load and fear of missing important visual details. The second theme, 'Preference for ADs: Dependent on the Video Content, Context, and BLV Individuals,' discusses the preference for pre-recorded ADs in immersive content and the differences in AD needs when watching alone versus socially. The third theme, 'User-driven AI: the Good and the Bad,' focuses on user-driven AI descriptions, noting that while they provide opportunity to describe a range of content, they sometimes lack context and suffer from hallucinations. AI text-to-speech voices are described as customizable.}
~\label{fig:themes}
\end{figure*}

We identified three themes based on the interviews and observation of participants' experiences with user-driven AI descriptions (Figure~\ref{fig:themes}). 

\subsubsection{\textbf{User-driven Description: Redefining Control for Accessibility}}

%Prior work on video accessibility has focused mainly on providing pre-determined audio descriptions within silent gaps in the video [cite]. However the video viewing landscape has changed drastically in the past few decades. Video viewing shifted from just on televisions to personal handheld devices, from movies to different forms of content such as short form content. Pre-determined inline or extended descriptions may not work for all video viewing scenarios, hence looking at different scenarios for different video accessibility formats is important. On demand is a novel approach for video accessibility which provides control to the BLV. 

The concept of control was mentioned repeatedly in participant descriptions of AD activations. The options to say when and how the descriptions were delivered created a sense of control that was missing from pre-recorded ADs. As P15$_{LB}$ highlighted: ``\textit{We're unlocking this absolutely mind-blowing layer that we can now add, which is an interactive audio description, the idea of having not only a description of what's going on, but a way to have the description describe to us what we want it to describe to us, how we want it to describe it to us.}'' P4$_{B}$ also echoed this sentiment, noting the limitations of pre-recorded ADs: ``\textit{When you have audio description built in, what is described is already predetermined... you can't really change that.}''. 
% Participants also noted that pre-recorded ADs often omit details due to timing constraints in a video: ``\textit{they [professional describers] just pick ... whatever they think, it is important. You miss a lot right?'' (P7)}
%being interactive and active

%use cases of on-demand AD (rewatching, different needs,
%\textbf{Use cases for on-demand ADs:} 
Participants enjoyed the flexibility of how and when they received ADs, giving them the ability to ``seek information'' (P4$_{B}$) exactly when they wanted it. For example, P6$_{B}$ activated descriptions based on whether they wanted to hear the person in the video or more detailed information about the visual content. The participants likened this user-driven description style to having a live ``human describer''.
%Several participants noted the experience to be active
% \textit{P7: I like the descriptions, and how I can activate them whenever I want. You know sometimes I prefer listening to the person in the video. Sometimes I wanted to hear more description. Sometimes I wanna hear more detailed information. So that was good to have that option.}
The ability to control ADs also helped cater to the needs of BLV individuals with varying visual impairments. For example, P15$_{LB}$ expressed their frustration with videos containing on-screen text, highlighting how he has to magnify or use OCR to read text; thus, he could get on-screen text in a more accessible manner using this interaction.
% For example, P11 noted that they preferred to rely on their functional vision, thus they could activate ADs to just read on-screen text when needed. % especially for the low vision participants, who requested descriptions to read on screen text. ``\textit{Where it's really helpful is the text, because reading text is something I just don't do when it comes to videos.}'' (P12). 
In contrast to pre-recorded ADs, the user-driven descriptions made the participants feel more active and engaged when watching videos. With this sense of control, P3$_{LV}$, P7$_{B}$, and P9$_{B}$ imagined creating and saving ADs that they or others could use to watch, summarize, or preview video content later. 

%BLV participants also noted specific scenarios when they wanted to control ADs. One example was action-packed videos when the professional describers had to omit details due to timing constraints in a video: ``\textit{they [professional describers] just pick ... whatever they think, it is important. You miss a lot right?'' (P7)} Others described the utility of this control when re-watching content: ``\textit{some of these Miyazaki films... there's all these different nuances to it...if you watch it again and you click, describe at different times, you're gonna understand all the foreshadowing pictures and things like that. So that part's a huge benefit.'' (P14)}. Similarly, P20 noted watching do-it-yourself content with concise ADs first, and requesting detailed descriptions when re-watching it later to find specific information.   the on-demand descriptions made the participants feel more active and engaged when watching videos. With this sense of control, P3 and P9 imagined creating and saving ADs with this on-demand approach that they or others could use to summarize or preview video content later.%Some participants mentioned using this for watching films to get all the visual details that they might miss out on in conventional AD, which P13 described as a potential hobby for BLV users. 

\textbf{Challenges of user-driven ADs.} The increased user control came with its challenges. One of the primary challenges was the disruption of video viewing experience and flow: ``\textit{It's easy to forget what was happening in the video if you use the on-demand descriptions too much}'' (P17$_{B}$). This disruption was more pronounced for content with greater visual details or little to no silent gaps.
% such as movies and fast-paced content with little to no silent gaps. 

% \textit{P19: Yeah, yeah. So you don't. Ideally, you know, it's like I'm following that audio. And then I request the description. I listen to the description. I would have forgotten what the person was kind of talking about. I mean, I would still remember the larger topic, but what exactly he was talking I would have forgot.}

The increased control associated with user-activated AD also led to a higher cognitive load. Participants had to actively engage with the video to know when to receive descriptions. This was mentally taxing, especially for content that was information-dense or had fewer silent gaps to activate descriptions. 
As P16$_{B}$ noted, ``\textit{I think the harder ones were things like the dog running through the course (Pets \& Animals). That's really hard because things are changing really fast... so it's just really challenging to keep up with that.}'' This cognitive load reduced utility for some participants, as they had to balance their attention between the content and activating AD. 

% \textit{P12: The only thing I noticed is, like, in that video, like you could use the, basically any videos that are fast when it comes to the action taking place. So, like the three I could think would be the fashion video, the dog show and the Kung Fu Panda trailer, the it's not that the audio descriptions are on point. I just found that it was a little bit tricky if you were going to depend on the descriptions entirely to maybe keep pace with some of the videos.}

User-driven ADs also instilled a fear of missing out (FOMO) visual information in some participants who expressed concern over missing critical information if they activated descriptions infrequently. This concern was particularly felt in fast-paced videos, where they could discern scene changes from audio but could not activate to keep pace with the scene changes:
% In other cases, the participants found it hard to discern scene changes, making it difficult to know when to activate descriptions:
``\textit{I was requesting descriptions too often, because I couldn't make those judgment calls.} (P7$_{B}$).'' Some participants felt they needed to press the keys at the right time or that they didn't press at the right moment to request the AD. This sense of FOMO could be a direct result of needing more time with user-driven ADs since it is a new way of receiving descriptions, as highlighted by five participants. 

\textbf{Strategies for activating ADs. } 
Participants had different strategies for activating descriptions. While some felt they activated descriptions randomly without any particular pattern ($n=3$), most noted that they relied on the audio track to decide when to activate an AD ($n=8$). They often avoided activating AD during conversations or speech and waited for a pause or silent gap in the video. Others noted activating descriptions when there were visual references (e.g., \textit{``do it like this}'' for the \emph{Health \& Fitness} video -- P18$_{B}$), when a change in the ambient sounds indicated a scene change (P19$_{B}$), or when they wanted to know the source of sounds in the audio track (e.g., \textit{``Chopping up something, okay what?}'' -- P13$_{B}$). User-driven ADs were particularly helpful for videos with difficult speech. For the \emph{Food and Cooking} video, even though the presenter explained all the steps, the speech was difficult to follow for some participants due to the presenter's accent, and the AD activations helped them understand the visual content better.

Participants also mentioned other strategies. Some noted that they requested descriptions early in the videos ``\textit{to get a context}'' (P9$_{B}$, P6$_{B}$, P20$_{B}$). Their interest in the content also affected when and how frequently they requested descriptions, with fewer and concise requests  %participants requested infrequently 
for videos that did not interest them. Finally, low vision participants also leveraged their functional vision to request ADs, especially to read on-screen text. ``\textit{Where it's really helpful is the text, because reading text is something I just don't do when it comes to videos.}'' (P10$_{LV}$).
% (add quote here from P11).
%requesting at random
% based on their interest in the content: Kung Fu Panda, requested very frequently
% for low vision, they used their functional vision to requesting when on-screen text
%Change in the audio track (ambient noise - P20)
%Using descriptions to complement info in audio track (sound of objects suggest sth is boiling, they request description to know it's noodles - P17)
%to know the cause or source of sounds (e.g., laughter - P9, P14
%avoided during conversations or waited for pauses in the video

% Descriptions were also requested early on in the video to get general context of the video (can add a visual related to this) or to reaffirm the visual content of the video.

% \textit{P10: I didn't know what the videos were going to be, so I was kind of curious to get something as early in the in the recording as I could to get a context of what what this was.}

\subsubsection{\textbf{Preference for ADs: Dependent on the Video Content, Context, and BLV Individuals.}} 
Participants noted that various forms of ADs (pre-recorded vs.\ user-driven, concise vs.\ detailed) supported different uses and that \textit{video content} and \textit{viewing context} played a crucial role in their preference for AD type. %The feedback highlighted how their preferences are determined by video content as well as viewing context for preference for concise or detailed AD, inline or on-demand format. 

{\bfseries Content dependency.} Conventional or pre-recorded AD format was preferred for entertainment content, where an immersive experience was essential ($n=6$). %These ADs are often created by professionals and integrated within silent gaps for minimal disruptions, creating a passive viewing experience. 
On the other hand, user-driven ADs were useful for content requiring active engagement over immersion, such as educational, workout, and cooking videos, where the ability to request ADs precisely when the action was happening enhanced their understanding of the content ($n=9$). 
The preference for concise vs.\ detailed ADs also depended on the video. For fast-paced videos or action sequences, concise descriptions were preferred to get sufficient information while minimally disrupting the video flow. %, and since they were shorter, made watching the video more efficient. 
Conversely, participants preferred detailed descriptions for heavily visual content (e.g., \emph{Film and Animation}), %. The detailed descriptions captured more visual detail about the background and characters, which helped with visualization 
as P9$_{B}$ highlighted: ``\textit{The detailed ones to me like they create a visual image in my mind more than you know short ones.}'' 
% This is also reflected in the statistics, where on average concise descriptions were requested much less than detailed requests for \emph{Film and Animation} (22.3\%). However, in the instructional (\emph{Food and Cooking}) video, concise was requested far more than detailed descriptions, 93.2\% more. 
For instructional videos, participants preferred to have detailed descriptions but noted that the detailed information was sometimes unrelated to the instructional aspect of the video and focused more on visual details, which reduced the utility of longer descriptions for instructional content as P2$_{B}$ highlights for the \textit{Cooking} video : ``\textit{I don't think you needed to know the details of everything... where they said the other stuff was in the kitchen like the food processor and the flowers, or whatever because that wasn't particular to the video.}'' 

%The preference for pre-recorded (inline or extended) AD vs on demand AD were also shaped by content and context dependency. Conventional AD format was preferred for entertainment purposes, where an immersive experience is essential. Typically these descriptions are created by professionals and well integrated within silent gaps of the content for minimal disruptions, creating a more passive viewing experience.
 
 {\bfseries Context dependency.} Participants' preference for the form of AD also depended on the viewing context, particularly whether they were watching or re-watching a video and whether they were viewing it alone or in a social setting (i.e., with sighted people). When asked if they would use user-driven descriptions in their daily life, participants indicated that their usage would vary depending on the situation. For instance, many found user-driven descriptions especially useful when re-watching videos. Sometimes, the purpose of re-watching was to get a new perspective or experience on the content.  ``\textit{Some of these Miyazaki films... there's all these different nuances to it...if you watch it again and you click, describe at different times, you're gonna understand all the foreshadowing pictures and things like that. So that part's a huge benefit.''} (P12$_{LB}$). In other cases, they wanted to re-watch the content for specific information. For instance, P19$_{B}$ noted watching do-it-yourself videos with concise ADs first and activating detailed descriptions when re-watching a part of the video later to find specific information. Participants also described different AD preferences when watching alone vs. with sighted people. P6$_{B}$ highlighted that in social contexts, ``\textit{the person who is watching with me, a sighted person will be so bored.}'' Hence, user-driven would be preferred more in a private setting rather than a social one. Similarly, P12$_{LB}$ noted that they would watch a video with frequent AD activation alone, then ``\textit{...watch it with other people and only click, describe occasionally when you're like, oh, I forgot what happened.}''

% \textit{
%         P9: I watch a lot of crops videos on YouTube, because I want to learn about crops, the gambling strategies. And so the visuals aren't important. It's just hearing the the the host like speak through the strategy, so I can hear exactly what you know what it is and hear him work through it. So I don't need description. I rarely watch videos that have no audio or no spoken content.
%     }

%Context in which the content is watched also impacted how participants felt they would use it in their daily life. P7 highlighted that in social contexts, \textit{"the person who is watching with me, a sighted person will be so bored."} Hence on demand would be preferred more in a private setting rather than a social one.

{\bfseries Individual differences in AD use and preference.}
The individual experiences of participants varied widely, which further highlights the importance of adaptable and customized ADs. Participants who appreciated having user-driven AD often cited it as a pleasant, enjoyable experience and a promising alternative to conventional AD (P1$_{B}$, P6$_{B}$, P14$_{B}$, P17$_{LB}$, P18$_{B}$). P6$_{B}$ noted: \textit{``I was trying to, you know, get as much as possible.''} However, not everyone shared the same narrative, and some found activating descriptions constantly to be overwhelming and unpleasant (P8$_{B}$, P10$_{LV}$, P20$_{B}$).
The varied and diverse opinions also translated into how participants activated descriptions, as shown in Figure~\ref{fig:examplerequests}. When asked if they would use user-activated ADs in daily video viewing, the preferred frequency of use varied widely among participants depending on their watching habits. Those who watched instructional and educational videos stated a higher frequency of use: ``\textit{it would be nice to kind of get the option...}''(P12$_{LB}$). 
Others, however, felt the disruption that comes with user-driven AD would not be suitable for the content they watch, particularly entertainment content. Hence why, participants felt that user-driven ADs would be more helpful for YouTube videos rather than movies or TV shows (P6$_{B}$, P15$_{LB}$, P19$_{B}$). Also, participants who mentioned a lower frequency of use felt that the content they watched was audio-based and did not require AD (P7$_{B}$, P8$_{B}$, P16$_{B}$).
% : ``\textit{[I] rarely watch videos that have no audio or no spoken content} (P9)'' 
%This is potentially a result of the inaccessibility of most video content. 
These participants also highlighted how they would use it for occasional descriptions: ``\textit{Seth Meyers in his show... a lot of his stuff has like silly graphics or things that they show. So when I hear him say something, a pause, and the audience like cracks up, I can run it and say, okay, there's a funny picture being shown of something, so that would be useful. (P8$_{B}$)}''

% \textit{
%         P13: I wasn't doing the descriptions for some of the ones that they were talking frequently. So like those ones I probably wouldn't go through. Not unless it's more instructional. Like the workout video one. I knew that knew the exercises that was she was doing, but wanted to just say, get a confirmation exactly which one she was doing, and stuff that most definitely would help, but like like that one for the Philippines again, that it's kind of more of an ad, I probably wouldn't have used it. But it is would be nice to kind of get the option, if I wanted to use it or not
%     }

\subsubsection{\bfseries{User-driven AI: the Good and the Bad}}
%Participants also shared feedback on the AI descriptions and the AI text to speech voice used in the study. 
Participants had varied opinions on the effectiveness of AI descriptions in the study and in general based on their prior experience and the types of content they watched. Several saw it as a promising future, while some felt skeptical about how well AI, even if user-driven, can describe different forms of content (P11$_{LB}$, P15$_{LB}$) . %AI misuse and uptake of AI for AD were highlighted as significant barriers while acknowledging the use of AI to reduce accessibility costs.
%We highlight the positive and negative aspects of using AI for video accessibility.

{\bfseries User-driven AI descriptions: the good.} %Participants appreciated how user-driven AI descriptions allowed them to address their unique preferences by the ability of choosing between concise and detailed descriptions: \textit{``the power of the technology will allow us to give people these choices to give them the concise or more flowery detail... because up to this point, you know, blind people have had very, very little described, and when we've had something described, we've had very little say in how it's described to us. (P9$_{B}$)''}. 
The user-driven approach provided an ``opportunity to describe more content''. The need for more content to be described was evident, as participants reported watching 11 unique genres, while traditional pre-recorded AD is often limited to film and TV shows. P19$_{B}$ found that user-driven AI descriptions will increase the range of content they can watch: \textit{``You know, like the [Legend of] Zelda, I would love to be able to sit there through a full description narration of game play-through.''} Additionally, AI descriptions with a user-driven approach can allow BLV users to efficiently browse diverse video content: \textit{``Just a quick description can let me know what's happening, and I can move on (P14$_{B}$).''} Relatedly, participants appreciated how user-driven AI allowed them to meet their unique preferences for the amount of detail: \textit{``the power of the technology will allow us to give people these choices to give them the concise or more flowery detail... because up to this point, you know, blind people have had very, very little described, and when we've had something described, we've had very little say in how it's described to us. (P9$_{B}$)''}.

Another significant benefit noted by participants was that user-driven AI AD can help eliminate subjective censorship by human describers. P1$_{B}$ mentioned how narrators are often too careful in describing people and characters to the point where they \textit{``lose some of the experience''}. A similar concern was echoed by P11$_{LV}$, who emphasized that descriptions should be on point and  \textit{``if you're embarrassed to narrate... I would hope you give your job to someone else.''} Ten participants appreciated the visual detail provided by the descriptions. By giving BLV users access to activate AI descriptions on demand, the user-driven approach can help mitigate concerns of subjectivity inherent in pre-recorded human descriptions. 

{\bfseries User-driven AI descriptions: the bad.} 
Participants highlighted several challenges with user-driven AI descriptions. Participants expressed that athough they provided visual detail, the descriptions lacked detailed guidance to follow along for instructional and educational content ($n=6$). This resulted in the AD activation providing more ``static'' rather than dynamic information, as noted by P16$_{B}$ in the \emph{Health and Fitness} video: \textit{``The motion was important to the content of the video... you have to know first you raise your right hand and your left leg, and that stuff is really really critical, and you can't capture that.''} On the other hand, AI descriptions sometimes assumed a certain context that was not provided. Even with a structured prompt, some descriptions referenced previous descriptions, which led to confusion as P7$_{B}$ stated that they \textit{``don't know what description 5 was''}. Although the prompt explicitly mentioned reading on-screen text central to understanding, the concise descriptions often failed to contain the on-screen text and only mentioned ``on-screen text.'' This further increased FOMO as expressed by P19$_{B}$: \textit{``What if one of the videos have a secret formula written on the screen, and only you know that the viewer... the sighted viewer could see that. But to the blind person, this just says there's text on the screen.''}

Participants mentioned AI hallucinations in some instances. The hallucinations were of particular concern for educational settings, as P13$_{B}$ explains:  \textit{``when things are slightly inaccurate, it shouldn't be used for things like exams... you know where it is not telling me the differences in the liquids, or like the video where it kept saying pork was chicken.''} 
Some participants explicitly noted concern with hallucinations in AI (P8$_{B}$). This was true even when AI descriptions were accurate. For example, in the comment for \emph{Pets \& Animal} video, P18$_{B}$ wrote:
\textit{``The dog part was mentioned towards the end of the video. However, the AD mentioned about the dog much before. This was helpful in getting the context into the video. Till the video spoke of dog, I was skeptical that the AI is calling something else a dog.''} These comments indicate a lower trust in AI-generated over human descriptions, which could hinder the adoption of a user-driven AI platform for several BLV participants (n=4).

Finally, while this approach provided agency over timing and detail, the descriptions did not adapt to the participants' backgrounds. While some participants enjoyed the visual \textit{``flowery detail (P6$_{B}$)''} provided by the descriptions, others highlighted how the descriptions implored too much background information. For instance, the mention of colors in the description made no sense to P16$_{B}$ as a congenitally blind person. These aspects highlight further expectations for flexibility and customization from AI descriptions. 

{\bfseries AI Text-to-Speech (TTS) voices: the good and bad.} The ability to adjust text-to-speech voices is particularly important for BLV users when it comes to user-driven interaction. The most common feedback from participants was wanting to speed up the TTS ($n=9$), noting that a faster TTS would enhance the efficiency of user-driven ADs. Several participants also highlighted the customization advantages of AI TTS, particularly the ability to change between male or female voices based on the content (P11$_{B}$, P15$_{LB}$). However, some participants noted AI voices can sound monotonous, lacking the emotional depth that a human describer would provide (P12$_{LB}$, P19$_{B}$), which was important for entertainment content for an immersive experience. For most other content, participants preferred a faster TTS to quickly browse the video. Overall, the quality and customization of AI descriptions was far more important to the participants than the TTS, \textit{``so long as the quality of the description itself is high, that's all that matters'' (P19$_{B}$).} 

\section{Discussion}

In this paper, we examined two questions: (Q1) What are BLV individuals' perceptions and experiences with user-driven AI-generated ADs? 
(Q2) How do BLV users' preferences for AD timing and detail differ between different video genres? 
Regarding Q1,  we identified the benefits and challenges associated with user-driven descriptions through detailed participant feedback. Participants appreciated the control and flexibility offered by user-driven interaction, but also highlighted the cognitive load and fear of missing out that it incurs. %AI descriptions aided visual understanding but lacked context. 
%User-driven descriptions offered participants a sense of flexibility, but at the cost of fear of missing information. 
Participants appreciated the opportunity to request descriptions for more diverse content, such as instructional videos and for re-watching content, and noted that user-driven descriptions might be less suitable for entertainment content, particularly when watching with others. There was also an interest for greater customization, such as adjusting the voice or tailoring descriptions to individual preferences.
For Q2, concise ADs were requested more often in our study, and the time interval between descriptions varied significantly across genres, with entertainment content requiring more frequent descriptions when compared to instructional content. These results provide insights into the utility of user-driven AI ADs and provide an avenue for further using the approach for a larger population of BLV users, as well as for diverse video content, to further assess emerging trends.  We discuss the need for an online user-driven AI description platform, the evolving role of describers and BLV users in AI-assisted AD authoring, and the potential of multisensory interactions in AD consumption.

\subsection{Implications for a User-Driven AI Description Platform}

A user-driven platform for AI-generated descriptions must offer customization of ADs, and various interactivity styles to meet the diverse needs of BLV users. In our study, BLV users watched various online content and wanted to use pre-recorded, user-driven, concise, and detailed ADs depending on the video content, social context, and their individual needs when watching a video. While some preferred having pre-recorded AD for entertainment content, others enjoyed having control over the ADs. Also, information that seemed trivial to some could be significant to others, such as the use of colors in descriptions. The diversity of BLV users' opinions and their patterns of AD activations underscores the need for customization in AI descriptions and AD systems. Relatedly, recent research have investigated various AD types and question-answering for BLV users~\cite{spica,yuksel2020HIML,automated_AD,ihorn2021narrationbot}. These AD approaches can be integrated into an online video-sharing platform (e.g., YouTube or YouDescribe) to provide agency to BLV users in watching diverse video content. With BLV users' permission, such an online platform can collect data on variations in video viewing preferences of BLV users (e.g., AD frequency, type, questions) over time to improve the AI descriptions and learn to time ADs depending on video content and user needs. This crowd-sourced approach can further enable new ways for BLV users to skim or re-watch videos or receive video descriptions.  

While user-driven AI descriptions present an opportunity for improving video accessibility, their real-time implementation introduces several practical challenges. In our study, we pre-generated descriptions to mitigate latency and provide a seamless viewing experience. However, generating descriptions live using LLMs can be time-consuming, especially for longer videos, and may lead to interruptions in the video viewing experience. One way to support user-driven interaction is to generate descriptions for all frames in advance, prior to the interaction. This introduces an initial delay in video preparation instead of during playback. While this delay may only take a few minutes for short videos, longer-form content could require significantly more time. Nonetheless, we anticipate that this latency will decrease as LLMs continue to evolve. Furthermore, to reduce cognitive load and prevent excessively long viewing sessions, these pre-generated descriptions can be dynamically grouped in real time into intervals based on the user's previous activations for the video, ensuring that essential information is not missed without requiring constant interaction. This approach minimizes the need for frequent activations while maintaining accessibility and reducing cognitive burden.

Nuances related to the accuracy and consistency of AI descriptions also need to be taken into account for AI-based AD platforms. Descriptions we generated using GPT-4-vision are high in accuracy, but a few descriptions still include object hallucinations (e.g., labeling ``pork ribs'' as ``chicken'' in the \emph{Food and Cooking} video)~\cite{Hallucinations}. 
% The hallucinations are especially hard to address in a user-driven platform when descriptions are generated live by an MLLM for each user. 
This highlights the importance of developing mechanisms to control the generation process and use human verification for AI descriptions when factual correctness cannot be determined via the audio track, or for critical information~\cite{gulfenvisioningcognitive}. In addition, MLLMs generate different outputs each time, raising questions about ensuring the quality of user-driven ADs. Contextual relevance can be another source of concern, especially for long-form content such as films and documentaries. Furthermore, even with prompt engineering, descriptions generated using MLLMs are prone to consistency issues. Although some inconsistencies might not notably affect the BLV user experience, issues such as self-reference to previous descriptions can increase the cognitive load for BLV users and reduce enjoyment. Future work can look into ways to ensure descriptions pass a quality threshold. For example, user-driven AD systems can include an automated evaluator module that checks the consistency of AI descriptions with guidelines. Also, a system that allows BLV users to provide feedback and co-create descriptions can help mitigate issues of accuracy, context, and consistency while improving the reliability of AI descriptions.

%We tested two levels of detail for the descriptions. Based on feedback, it is obvious that BLV people's needs and preferences encompass the two sets of details provided. Information that may seem trivial to some could be significant to others, such as the use of colors in descriptions. Future work should explore providing user-driven descriptions with various customizations for BLV users.

%BLV preferences need to be accounted for when creating AI AD platforms. This means taking into account different options on how to receive AD. As highlighted in the feedback, some participants preferred having pre-determined AD over user-driven AD, especially for entertainment content. While others enjoyed user-driven AD, the level of detail did not always provide visual information that BLV participants wanted. To cater to this, a platform should also provide the ability to ask questions at any given point, and have the ability for the user to choose between having pre-determined AD or a user-driven description interaction or having both.

\subsection{Change in Roles for Describers and BLV Users}
With recent advances in AI descriptions, there is a potential for the roles of novice describers to evolve from creating descriptions to adjusting the timings of receiving descriptions and removing hallucinations and inaccuracies. Although we did not recruit any sighted describers for our study, our results suggest new ways for novice describers to support BLV user needs. %how their role could evolve based on previous work. 
%Prior work has explored how preset sentence templates can enhance descriptions generated by novice describers for images~\cite{image_templates}. Similarly, creating descriptions with automated feedback~\cite{natalie_feedback} and automating video text generation and scene segmentation~\cite{yuksel2020HIML} was easier for novice describers and significantly improved the AD quality for BLV users. 
While previous efforts have mainly focused on using AI to support novice describers in creating or editing ADs~\cite{image_templates,natalie_feedback,yuksel2020HIML}, % enhance description quality, and reduce the cost of audio description creation, 
with newer MLLMs, the role of describers could shift. In terms of providing visual details, most BLV participants felt the AI descriptions were adequate. However, several participants needed help with knowing when to request descriptions. Instead of describers working to create descriptions, they can work on when to insert descriptions. Although libraries and models exist that help detect silent gaps in audio for the insertion of descriptions, these fail for short-form content where there are fewer or no silent gaps. Some BLV users also had low trust in AI-generated descriptions, which worsened with even small AD inaccuracies. Sighted describers can easily verify the accuracy or detect such hallucinations and inconsistencies and correct them to improve BLV user comprehension, trust, and overall experience.
%How about automated timing with scene changes

With user-driven AI descriptions, the role of BLV users can also further shift from video consumption to active content creation. Some BLV users in our study were excited about the possibilities of user-driven ADs and wanted to save and share ADs (and variations on every re-watching) for the videos. Also, %several use cases emerged when BLV users requested descriptions (e.g., on-screen text, visual reference). 
compared to sighted novice describers, low vision volunteers have a better idea of when descriptions are needed for a video (e.g., on-screen text, visual reference). While there are already BLV content creators~\cite{digitalContentCreationBLV, avscript}, a user-driven AI tool can open up the space to a wider range of BLV users to become AD creators, determining when and how ADs should be inserted for different videos based on their lived experience. Specifically, AD tools can incorporate the user-driven approach with additional functionalities for BLV users to review, edit, save, and share ADs to enable more than momentary access to descriptions and support BLV people as curators of content that goes beyond their own personal consumption. This role change can make AD creation an inclusive space for BLV describers.
% AD creation is a collaborative effort, discuss focus groups and link it back to how this inclusion can be helpful

\subsection{User-Driven AD with Multisensory Interactions}
Prior research has looked into using multi-sensory AD approaches for artwork~\cite{multiADVisualArts, accessibleVisualArts, artVoiceControlled} and for AD in movies ~\cite{hapticsInMovies, midairhaptics} to improve visualization. BLV studies have also highlighted the need for additional output modalities such as audio cues, tactile graphics, and haptics to enhance the AD experience for different viewing scenarios (e.g., how-to, short-form, comedy, drama)~\cite{jiang2024s} and support user agency in navigating 360$^\circ$ videos~\cite{jiang360videos}. 
Though this requires technology that might not always be available to BLV users, several output modalities can be incorporated for ADs. Our participants wished to know when to activate AD for user-driven style AD. Brief audio cues, or vibrations, can be incorporated into the video to let the BLV user know when they can activate a description.
By adding audio or haptic cues, the control of receiving descriptions still lies with the user, and they have a better idea of when there are scene changes or where they can benefit from activating descriptions, effectively reducing the cognitive load of not knowing when to request descriptions or having too many descriptions in the case of pre-recorded AD. In this regard, MLLMs can help in providing this multi-sensory information. For example, AI can detect scene changes, on-screen text, character changes, visual references, and generate corresponding audio cues and vibrations. The frequency of such feedback can be tailored based on user preferences and customizations. 
Currently, limited work has been done on the intersection of AD and other modalities, especially how BLV users engage with personal devices (i.e., smartphones) for video consumption~\cite{accessibilityresearch}.
% Similarly, few studies examine how BLV users engage with smartphones for accessible video media~\cite{accessibilityresearch}. 
Future work can explore whether other modalities complement or obstruct BLV users' attention to AD.

\subsection{Limitations and Future Work}

Our work has several limitations.
First, our study covers a subset of video types that BLV users wish to watch. Within a 75-minute study, we could only test seven video genres, with one short video per genre. 
% compared to the fifteen video genres available on YouDescribe. 
% Also, all the videos were relatively short. 
The limited selection restricts the generalizability of our findings, as a lot of variation exists in content within each genre.  The pace of the video likely impacted the quantitative results. Testing user-driven AD with other genres, including a broader range of videos with different paces and longer durations, can give further insight into the efficacy of user-driven descriptions across different video types.
Second, we created all the descriptions prior to the user study to account for latency and description quality. Pre-generating descriptions allowed us to reduce time delays and control for the effect of variations in AI generation on BLV user ratings. This approach also enabled us to process any formatting issues in the generated descriptions. A user-driven AD platform must include automated methods to mitigate these issues in the descriptions generated on the spot. Third, we did not evaluate whether the generated descriptions fully adhered to the curated guidelines. Our study was based on prior work, which has shown that prompting MLLMs with AD guidelines improved description quality in terms of clarity, accuracy, objectivity, and descriptiveness compared to descriptions generated by either baseline MLLMs or novice describers~\cite{li2025videoa11y}. However, future work should devise mechanisms to measure how well user-driven descriptions comply with the guidelines.
Fourth, we investigated BLV people's perception of user-driven ADs in a single session.  %our user study was conducted with participants with prior experience with AD and good computer literacy. 
Some participants highlighted they would require more time with this new way of watching videos, and their frequency of use might differ for the content they watch on a regular basis. To gather more information on user-driven AD, a longitudinal study in a more organic setting needs to be conducted. This would provide further insights into how BLV users' interest in the video content and their feelings about user-driven ADs can change over a more extended period and how these factors can impact the frequency of requests for description. Lastly, the study was conducted with 20 BLV participants. While this sample provided rich qualitative insights into BLV perceptions of user-driven AI descriptions, the number of participants limits the generalizability of our statistical results. Future work could recruit a larger number of participants to further expand on our findings.

\section{Conclusion}
We presented an alternative approach to video accessibility by giving more control to the BLV individuals through user-driven descriptions with two types of detail. Based on the study results with 20 BLV participants, user-driven descriptions improved the sense of user control with flexibility of when and how participants receive descriptions but increased cognitive load and fear of missing out (FOMO). BLV users found the descriptions compelling in terms of the detail of visual content and saw a possibility to engage with more types of content through user-driven interaction. There were also concerns about AI's ability to describe certain kinds of genres. %Participants also raised context relevance as a significant concern. 
%User-driven AD instilled a fear of missing out on crucial visual content as it was often difficult to discern different scene changes. Participants highlighted they would benefit from having some way of knowing when to request descriptions. Additionally, user-driven was of more utility for certain video genres (e.g., instructional, educational), while pre-determined AD was preferred for others (e.g., film and animation). 
With the rapid advances in AI, we hope our results can inform future work on tailoring AI descriptions based on user preferences and customization needs and open up further possibilities for BLV users to access and create AD content based on their needs and lived experiences.

\begin{acks}
 This research was supported by the National Eye Institute (NEI) of the National Institutes of Health (NIH) under award number R01EY034562. The content is solely the responsibility of the authors and does not necessarily represent the official views of the NIH. 
\end{acks}

\bibliographystyle{ACM-Reference-Format}
\bibliography{myrefs}
% \nocite{*}

\newpage
\appendix
%TC:ignore

\section{Audio Descriptions Guidelines}
\label{app:ad}
The list below shows the complete 42 audio description (AD) guidelines we curated for prompting GPT-4 Vision (GPT-4V) to create descriptions. Each guideline is listed alongside its corresponding source. For our purposes, we adapted the original guidelines by shortening and combining related guidelines for brevity.

\setlength{\fboxsep}{8pt} 
\setlength{\fboxrule}{0.5mm}
\vspace{0.2cm}
\noindent
\fcolorbox{black}{blue!2}{ 
    \begin{minipage}{0.94\linewidth}
    \raggedright
        1. Avoid over-describing — Do not include non-essential visual details.~\cite{netflix_style_guide}
        
        2. Description should not be opinionated unless content demands it.~\cite{netflix_style_guide}
        
        3. Choose level of detail based on plot relevance when describing scenes.~\cite{netflix_style_guide}
        
        4. Description should be informative and conversational, in present tense and third-person omniscient.~\cite{netflix_style_guide}
        
        5. The vocabulary should reflect the predominant language/accent of the program and should be consistent with the genre and tone of the content while also mindful of the target audience. Vocabulary used should ensure accuracy, clarity, and conciseness.~\cite{netflix_style_guide}
        
        6. Consider historical context and avoid words with negative connotations or bias.~\cite{netflix_style_guide}
        
        7. Pay attention to verbs — Choose vivid verbs over bland ones with adverbs.~\cite{netflix_style_guide}
        
        8. Use pronouns only when clear whom they refer to.~\cite{netflix_style_guide}
        
        9. Use comparisons for shapes and sizes with familiar and globally relevant objects.~\cite{netflix_style_guide}
        
        10. Maintain consistency in word choice, character qualities, and visual elements for all audio descriptions.~\cite{netflix_style_guide}
        
        11. Tone and vocabulary should match the target audience's age range.~\cite{netflix_style_guide}
        
        12. Ensure no errors in word selection, pronunciation, diction, or enunciation.~\cite{dcmp_description_key}
        
        13. Start with general context, then add details.~\cite{dcmp_description_key}
        
        14. Describe shape, size, texture, or color as appropriate to the content.~\cite{dcmp_description_key}
        
        15. Use first-person narrative for engagement if required to engage the audience.~\cite{dcmp_description_key}
        
        16. Use articles appropriately to introduce or refer to subjects.~\cite{dcmp_description_key}
        
        17. Prefer formal speech over colloquialisms, except where appropriate.~\cite{dcmp_description_key}
        
        18. When introducing new terms, objects, or actions, label them first, and then follow with the definitions.~\cite{dcmp_description_key}
        
        19. Describe objectively without personal interpretation or comment. Also, do not censor content.~\cite{dcmp_description_key, ofcom_access_services}
        
        20. Deliver narration steadily and impersonally (but not monotonously), matching the program's tone.~\cite{ofcom_access_services}
        
        21. It can be important to add emotion, excitement, lightness of touch at different points. Adjust style for emotion and mood according to the program's genre.~\cite{ofcom_access_services}
        
        22. If it is children’s content, tailor language and pace for children's TV, considering audience feedback.~\cite{ofcom_access_services}        
    \end{minipage}
}

\setlength{\fboxsep}{8pt} 
\setlength{\fboxrule}{0.5mm}
\noindent
\fcolorbox{black}{blue!2}{ 
    \begin{minipage}{0.94\linewidth}
    \raggedright              
        23. Do not alter, filter, or exclude content. You should describe what you see. Try to seek simplicity and succinctness in your description.~\cite{mediaccess}
        
        24. Prioritize what is relevant when describing action as to not affect user experience.~\cite{netflix_style_guide}
        
        25. Include location, time, and weather conditions when relevant to the scene or plot.~\cite{netflix_style_guide}
        
        26. Focus on key content for learning and enjoyment when creating audio descriptions. This is so that the intention of the program is conveyed.~\cite{dcmp_description_key}
        
        27. When describing an instructional video/content, describe the sequence of activities first.~\cite{dcmp_description_key}
        
        28. For a dramatic production, include elements such as style, setting, focus, period, dress, facial features, objects, and aesthetics.~\cite{dcmp_description_key}

         29. Describe what is most essential for the viewer to know in order to follow, understand, and appreciate the intended learning outcomes of the video/content.~\cite{dcmp_description_key}

        30. Audio description should describe characters, locations, time and circumstances, on-screen action, and on-screen information.~\cite{ofcom_access_services}
        
        31. Describe only what a sighted viewer can see.~\cite{mediaccess}
        
        32. Describe main and key supporting characters' visual aspects relevant to identity and personality. Prioritize factual descriptions of traits like hair, skin, eyes, build, height, age, and visible disabilities. Ensure consistency and avoid singling out characters for specific traits. Use person-first language.~\cite{netflix_style_guide}
        
        33. If unable to confirm or if not established in the plot, do not guess or assume racial, ethnic or gender identity.~\cite{netflix_style_guide}
        
        34. When naming characters for the first time, aim to include a descriptor before the name (e.g., a bearded man, Jack).~\cite{netflix_style_guide}
        
        35. Description should convey facial expressions, body language and reactions.~\cite{netflix_style_guide}
        
        36. When important to the meaning / intent of content, describe race using currently-accepted terminology.~\cite{dcmp_description_key}
        
        37. Avoid identifying characters solely by gender expression unless it offers unique insights not apparent otherwise to visually impaired viewers.~\cite{mediaccess}
        
        38. Describe character clothing if it enhances characterization, plot, setting, or genre enjoyment.~\cite{mediaccess}
        
        39. If text on the screen is central to understanding, establish a pattern of on-screen words being read. This may include making an announcement, such as 'Words appear'.~\cite{mediaccess}
        
        40. In the case of subtitles, the describer should read the translation after stating that a subtitle appears.~\cite{mediaccess}
        
        41. When shot changes are critical to the understanding of the scene, indicate them by describing where the action is or where characters are present in the new shot.~\cite{netflix_style_guide}
        
        42. Provide description before the content rather than after. ~\cite{netflix_style_guide}
    \end{minipage}
}

\clearpage

\section{GPT-4V description prompt}
\label{app:prompt}
The following prompt was used to create concise and detailed description using the GPT-4V API.

\setlength{\fboxsep}{8pt} 
\setlength{\fboxrule}{0.5mm}
\vspace{0.2cm}
\noindent
\fcolorbox{black}{blue!2}{ 
    \begin{minipage}{0.94\linewidth}
    \raggedright
    Imagine you are a video description expert. You will watch some keyframes from a video. For each keyframe, you will provide video descriptions, following the guidelines below. The instructions are some of the guidelines to keep in mind when providing descriptions to blind and low vision users.
    
    You'll need to use [all] of the guidelines.
    
    [42 Audio Description Guidelines]
    
    Once you have created the description following the guidelines, revisit the guidelines to see of anything is missing and edit the description to closely adhere to the guidelines.
    
    As these descriptions will be spoken for the second of the video they describe, the descriptions should sound as if they are part of a continuous narrative rather than segmented into distinct frames or images. Do not mention frames in your description.
    
    Return each description in a new line. Each audio description should be around [maximum word limit] words. If you cannot provide descriptions, provide a reason as to why.
    \end{minipage}
}

\section{Descriptions}
\label{app:descriptions}
The following figures showcase a frame and its corresponding concise and detailed descriptions for each of the seven videos used in the study. The concise descriptions often omit the on-screen text and include less information and adjectives about the visual scene, objects, actions, and people's appearances compared to the detailed descriptions. The hallucinations in the descriptions are highlighted with \textcolor{red}{\underline{underlined red}} text.

\begin{figure}[htbp]
    \centering
    \fcolorbox{black}{blue!2}{%
        \begin{minipage}{0.94\columnwidth}
            \includegraphics[width=\columnwidth]{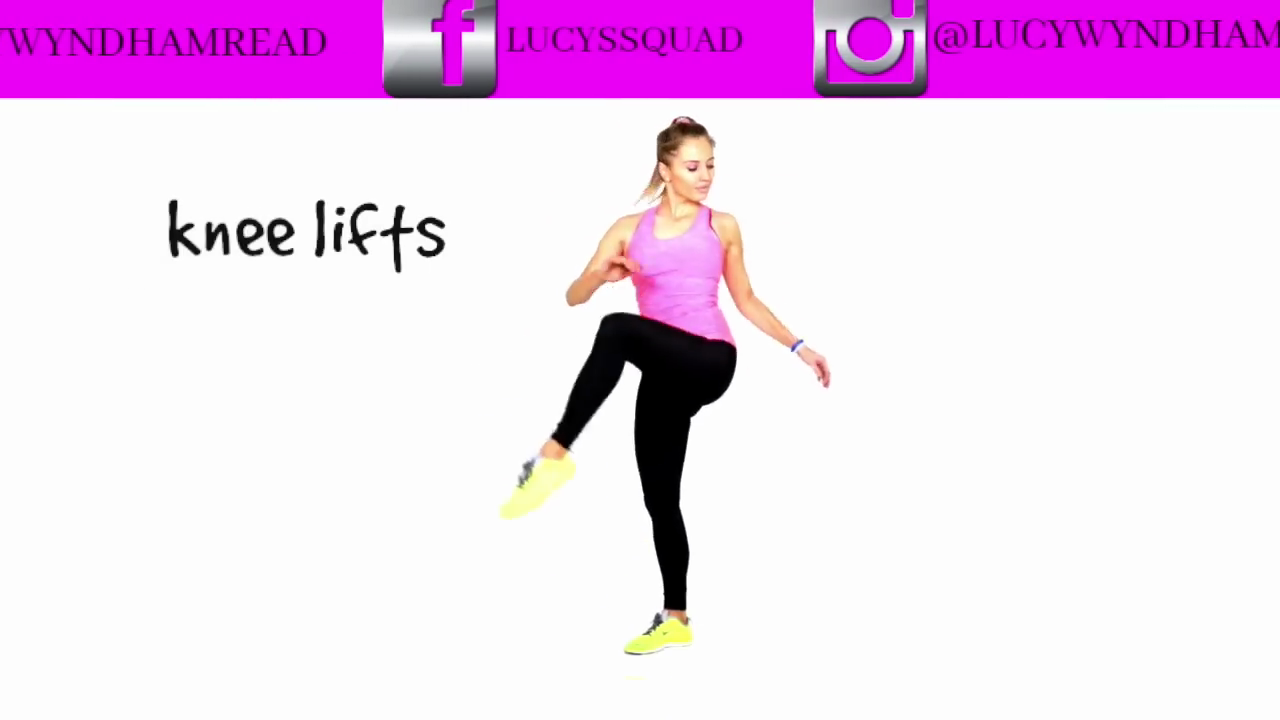}
            \Description{A frame from an exercise video. A fitness instructor is performing knee lifts. She raises one knee while touching it with the opposite hand. The text ``knee lifts'' is displayed above her in the video. The background is white with social media handles at the top.}
            \captionof{figure}{Health \& Fitness Video} 
            \textbf{Concise Descriptions:} \\
            The instructor transitions to knee lifts, bringing up one knee while the opposite hand touches it. \\

            \textbf{Detailed Descriptions:} \\
            Now, the focus shifts as the woman begins knee lifts. \textit{"knee lifts"} is displayed in text above her. She raises her right knee and brings her hands towards the lifted knee.
        \end{minipage}%
    }
\end{figure}

\begin{figure}[htbp]
    \centering
    \fcolorbox{black}{blue!2}{%
        \begin{minipage}{0.94\columnwidth}
            \includegraphics[width=\columnwidth]{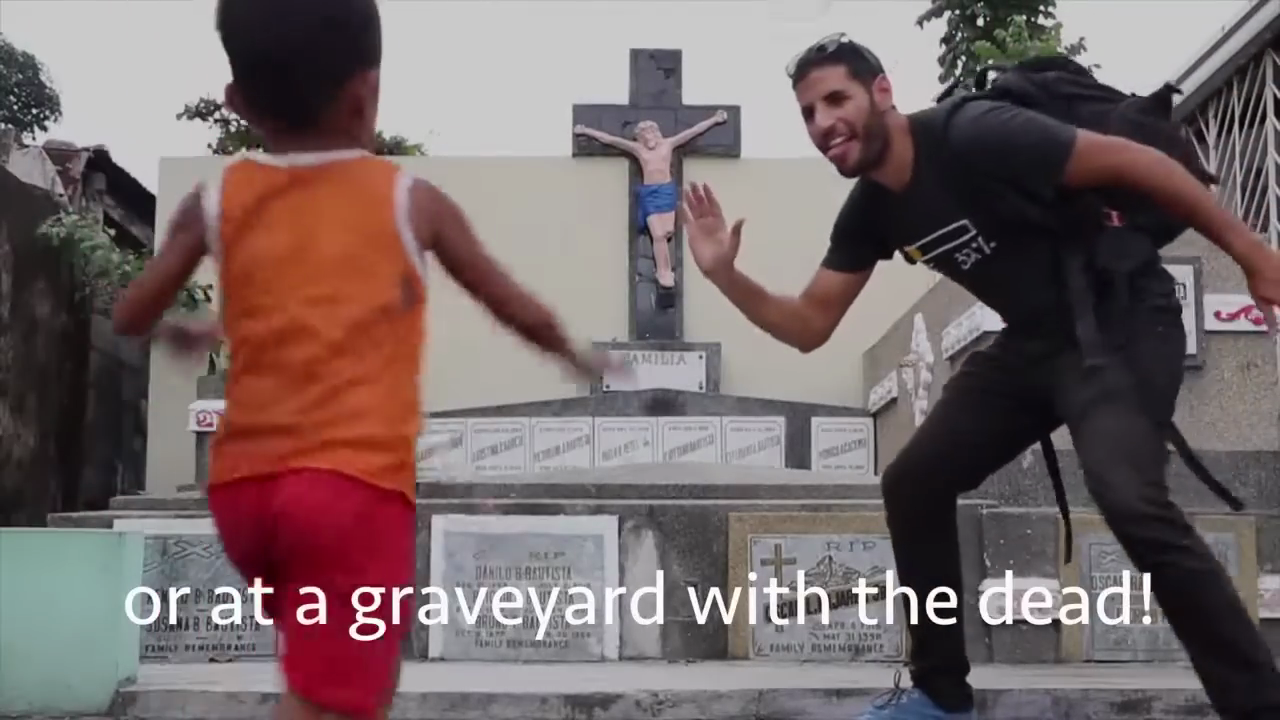}
            \Description{People and Blogs video shows a frame from a graveyard setting with crosses and headstones. The overlay text reads, ``or at a graveyard with the dead!''}
                \captionof{figure}{People and Blogs Video} 
                \textbf{Concise Descriptions:} \\
                A man playfully interacts with a child in front of crosses at a graveyard, evoking a mix of life and death.\\

                \textbf{Detailed Descriptions:} \\
                A man interacts animatedly with a young child near a wall adorned with religious symbols and graves, hinting at a community with unique living circumstances. The subtitle suggests an unusual coexistence with the deceased. \\
        \end{minipage}%
    }
\end{figure}

\begin{figure}[htbp]
    \centering
    \fcolorbox{black}{blue!2}{%
        \begin{minipage}{0.94\columnwidth}
            \includegraphics[width=\columnwidth]{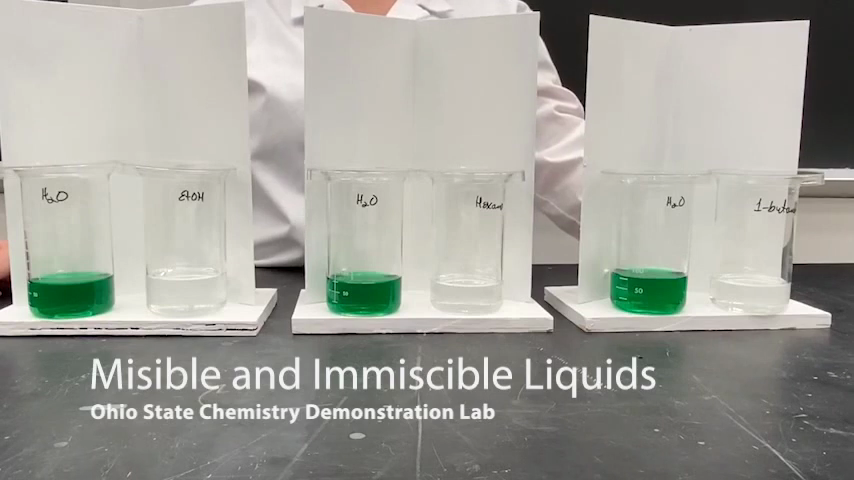}
            \Description{An educational video showing a chemistry demonstration of beakers. The text ``Miscible and Immiscible Liquids'' is displayed, along with the label ``Ohio State Chemistry Demonstration Lab.''}
                \captionof{figure}{Education Video} 
                \textbf{Concise Descriptions:} \\
                \textcolor{red}{\underline{Three beakers}} on a lab bench, labeled H2O, EtOH, and 1-butanol, each containing a green liquid, demonstrating ``Miscible and Immiscible Liquids.''\\

                \textbf{Detailed Descriptions:} \\
                An over-the-counter view reveals \textcolor{red}{\underline{three beakers}} against a white backdrop, with ``H2O'' labeled on the first and third beaker, ``EtOH'' on the second, and ``1-butanol'' on the \textcolor{red}{\underline{fourth}}. A green liquid rests at the bottom of each beaker, and text above reads, ``Miscible and Immiscible Liquids. \textcolor{red}{\underline{Otto}} State Chemistry Demonstration Lab'' \\
        \end{minipage}%
    }
\end{figure}

\begin{figure}[htbp]
    \centering
    \fcolorbox{black}{blue!2}{%
        \begin{minipage}{0.94\columnwidth}
            \includegraphics[width=\columnwidth]{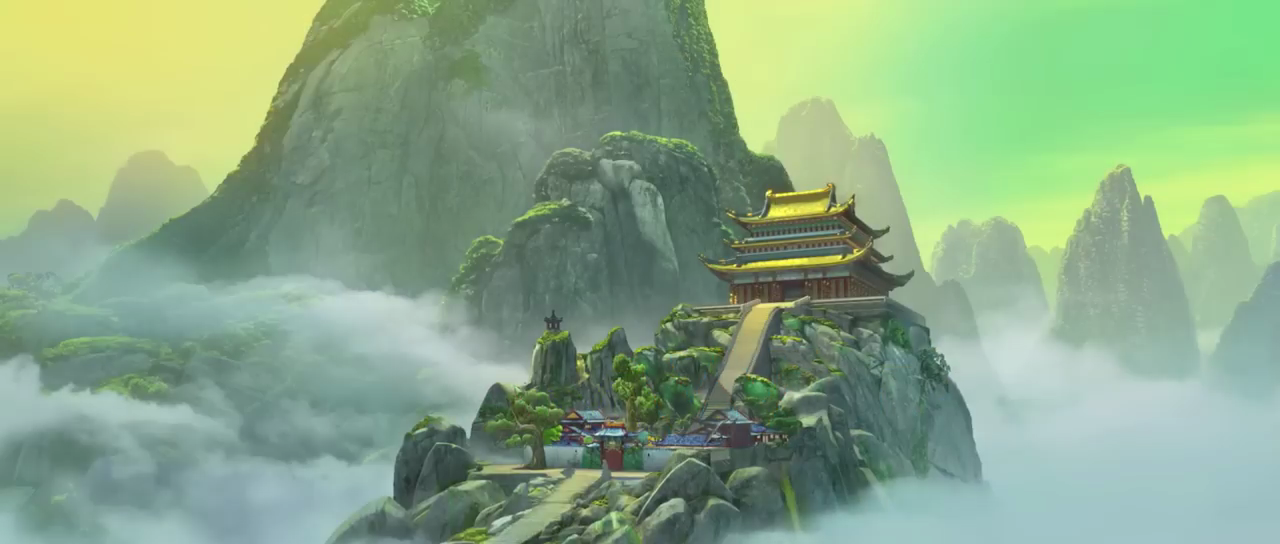}
            \Description{A frame from Kung Fu Panda trailer displaying an animated, mystical landscape with a pagoda surrounded by swirling mist. The mountain peaks tower against a golden sky.}
                \captionof{figure}{Film and Animation Video}  
                \textbf{Concise Descriptions:} \\
                An animated, mystical mountain landscape with a pagoda, surrounded by clouds and shrouded in a soft, golden light. \\

                \textbf{Detailed Descriptions:} \\
                A panoramic vista shows an ethereal mountaintop temple amidst towering peaks, with swirling mists and a glowing, golden sky accentuating the serene setting.
        \end{minipage}%
    }
\end{figure}

\begin{figure}[htbp]
    \centering
    \fcolorbox{black}{blue!2}{%
        \begin{minipage}{0.94\columnwidth}
            \includegraphics[width=\columnwidth]
            {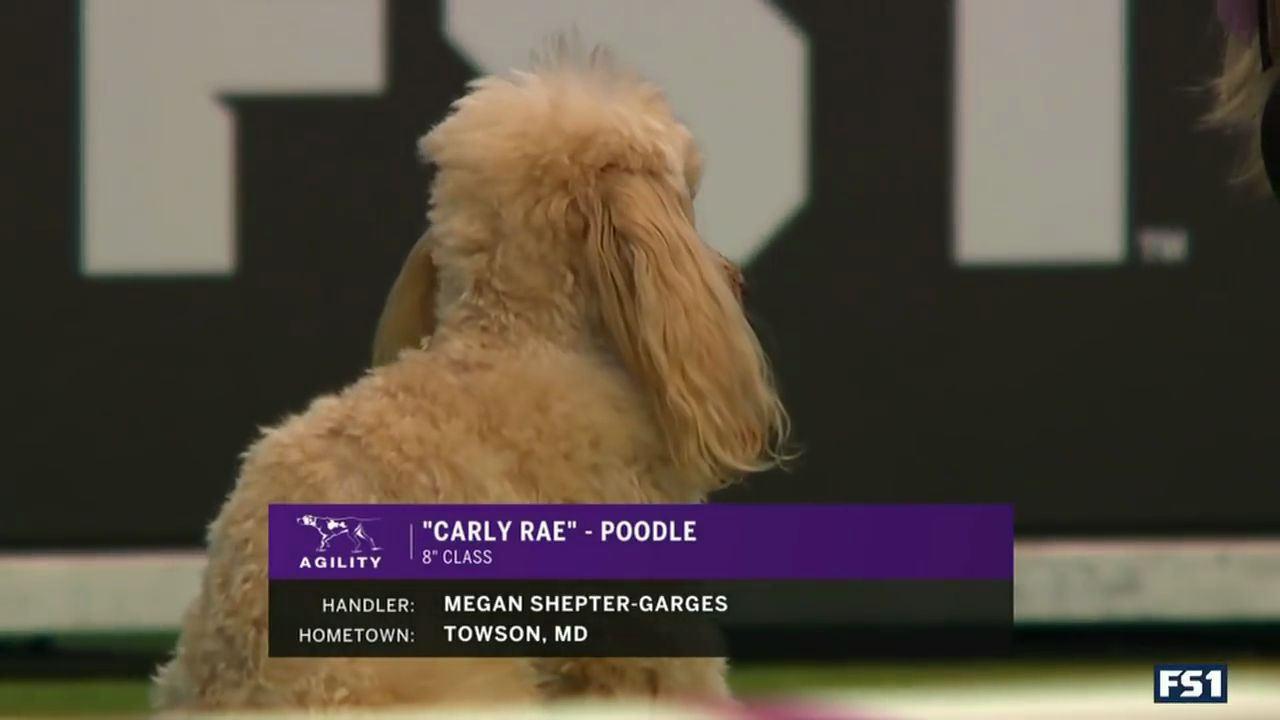}
            \Description{A frame from the Pets & Animals video showing a light brown poodle named "Carly Rae" from behind during an agility competition. The purple banner on the screen displays details: "Carly Rae - Poodle, 8" Class." Below, the handler is identified as Megan Shepter-Garges from Towson, MD. The FS1 logo is visible in the bottom right corner.}
                \captionof{figure}{Pets \& Animals}
                \textbf{Concise Descriptions:} \\
                A caramel-colored poodle named "Carly Rae" is seen with text detailing its class, handler, and hometown.\\

                \textbf{Detailed Descriptions:} \\
                A fluffy apricot poodle stands attentively in an indoor setting with a purple and white graphic displaying the name "CARLY RAE," indicating the dog's class as "8" and listing handler and hometown details. \\
        \end{minipage}%
    }
    
\end{figure}

\begin{figure}[htbp]
    \centering
    \fcolorbox{black}{blue!2}{%
        \begin{minipage}{0.94\columnwidth}
            \begin{center}
            \includegraphics[height=3cm]         
            {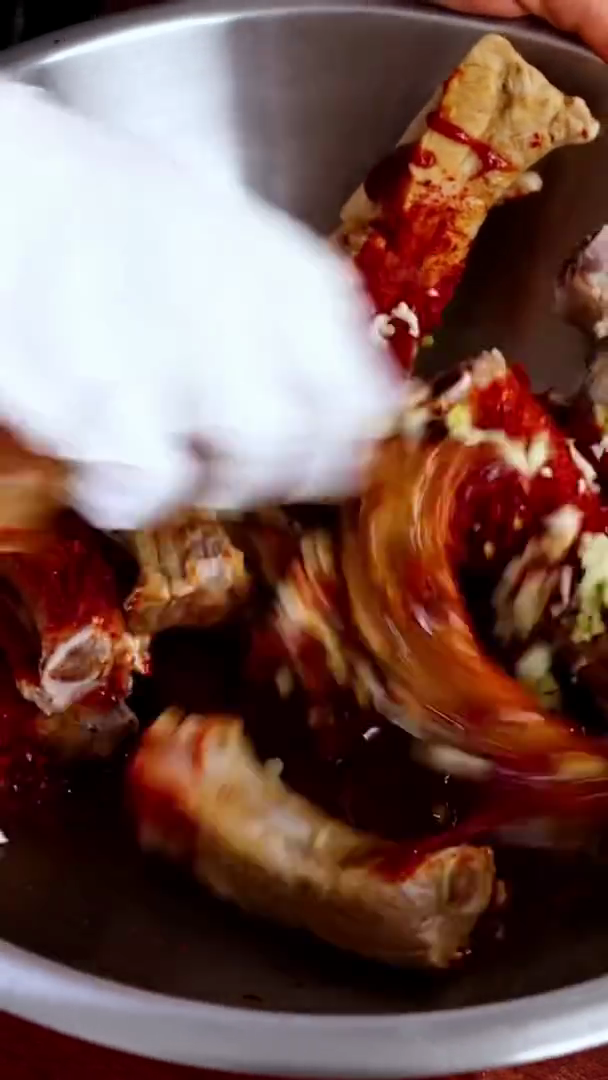}
                \captionof{figure}{Food \& Cooking Video}
                \Description{ A frame from the cooking video displays a hand wearing a white glove and tossing pork ribs in a metallic bowl filled with sauce, coating them evenly.}
            \end{center}
                \textbf{Concise Descriptions:} \\
                Gloved hands toss \textcolor{red}{\underline{chicken wings}} in a bowl with a spicy red sauce. \\

                \textbf{Detailed Descriptions:} \\
                A hand in a white glove mixes seasoned \textcolor{red}{\underline{chicken wings}} in a metallic bowl, coating them with a red, thick, sticky sauce.
        \end{minipage}%
    }
\end{figure}

\begin{figure}[htbp]
    \centering
    \fcolorbox{black}{blue!2}{%
        \begin{minipage}{0.94\columnwidth}
            \begin{center}
            \includegraphics[height=3cm]         
                {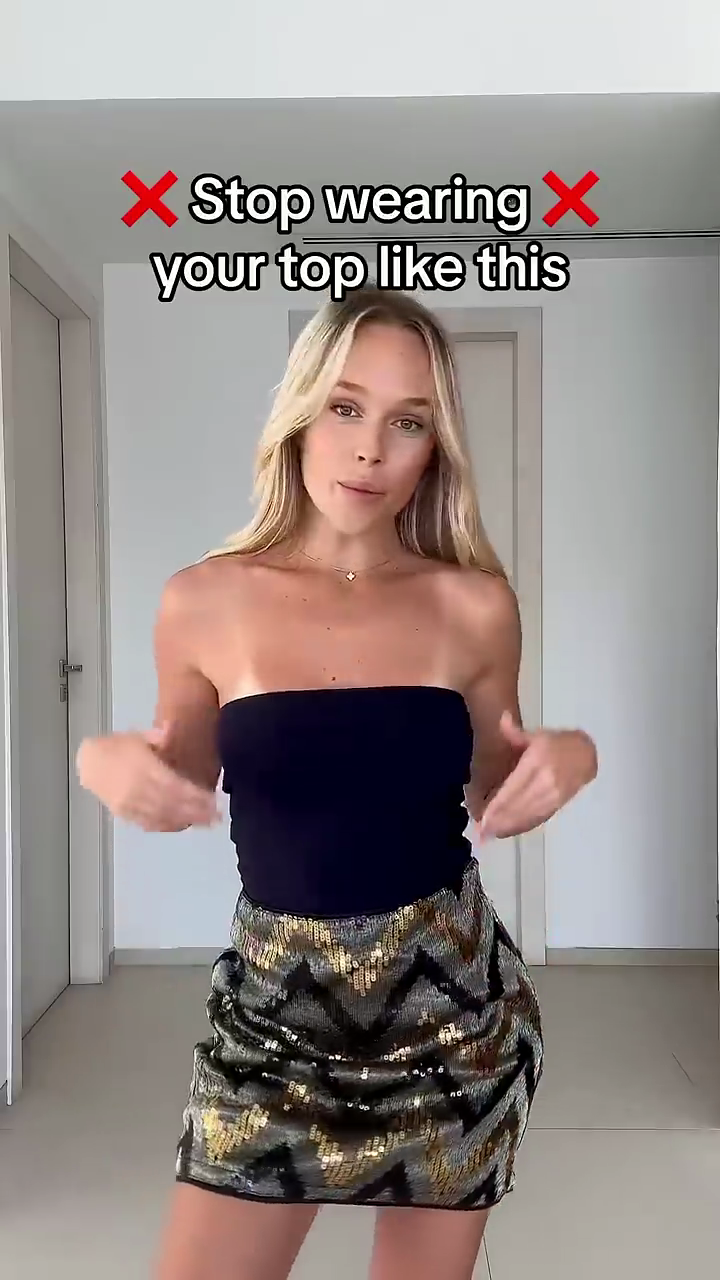}
                \captionof{figure}{Beauty Video} 
                \Description{A beauty video showing a woman wearing a strapless black top and sequined skirt. The text above her reads, "Stop overthinking your outfit choices," with a red "X".}
            \end{center}
                \textbf{Concise Descriptions:} \\
                A woman stands in a room, wearing a strapless top and sequined skirt, a red X above her head. \\

                \textbf{Detailed Descriptions:} \\
                A woman stands in a brightly lit room, with a displeased expression, wearing a strapless black top and a sequined skirt. Red crosses and the words ``Stop wearing your top like this’’ emphasize her message.
        \end{minipage}%
    }
\end{figure}

\clearpage
\section{Appendix: Interview Questions in Our User Study}
\label{app:interview}
The following questions were used during the post-session interviews to gather participants’ qualitative feedback on their experience with user-driven interaction. The interview was semi-structured, allowing for follow-up questions based on the responses.

\setlength{\fboxsep}{8pt} 
\setlength{\fboxrule}{0.5mm}
\vspace{0.2cm}
\noindent
\fcolorbox{black}{blue!2}{ 
    \begin{minipage}{0.97\linewidth}
    \raggedright

\begin{itemize}
    \item How was your experience with user-driven descriptions?
    \item How was the experience of user-driven descriptions as compared to your previous experience with AD?
    \item How did you decide whether to activate a description (i.e., what factors impacted your decision)?
    \item What do you think is the best way to receive descriptions?
    % : user-driven approach or at predetermined times?
    % \item Follow-up: What determined your preference?
    \item I noticed you requested descriptions (frequently/rarely) for (most/some) videos. Was there a reason for that?
    \item If you could have user-driven descriptions for any video, would you have a similar frequency of activations in daily life?
    \item What did you think about the AI descriptions?
    \item How do you think the AI descriptions compare to descriptions created by human describers?
    \item How was your experience of AI voice being used for descriptions?
    \item What are your thoughts about the concise descriptions?
    \item What are your thoughts about the detailed descriptions?
    \item In what situations would you prefer concise descriptions? In what scenarios would you prefer detailed descriptions?
    \item Were there any videos for which you felt the descriptions were good? 
    \item Were there any videos in which you felt the descriptions were bad?
    \item How could the descriptions be made better for the videos with bad descriptions?
    \item In the end, do you have any other comments?
\end{itemize}

    \end{minipage}
}

\end{document}